\documentclass[preprintnumbers,floatfix,superscriptaddress,nofootinbib]{revtex4-1}

\usepackage{amsmath,amssymb}
\usepackage{graphicx}
\usepackage[dvipsnames]{xcolor}
\usepackage[linktocpage=true,colorlinks=true,urlcolor=brown,linkcolor=blue,citecolor=magenta]{hyperref}
\usepackage[english]{babel}
\usepackage{cancel}
\usepackage{parskip}
\usepackage[normalem]{ulem}

\newcommand{\Sec}[1]{Sec.~\ref{#1}}

\newcommand{\Fig}[1]{Fig.~\ref{#1}}

\newcommand{\Eq}[1]{Eq.~(\ref{#1})}
\newcommand{\Eqs}[2]{Eqs.~(\ref{#1}) and (\ref{#2})}

\newcommand{\eg}{\textit{e.g.}}
\newcommand{\ie}{\textit{i.e.}}

\newcommand{\beq}{\begin{equation}}
\newcommand{\eeq}{\end{equation}}
\newcommand{\ba}{\begin{array}}
\newcommand{\ea}{\end{array}}
\newcommand{\bea}{\begin{eqnarray}}
\newcommand{\eea}{\end{eqnarray} }
\newcommand{\be}{\begin{eqnarray}}
\newcommand{\ee}{\end{eqnarray}}
\newcommand{\bal}{\begin{align}}
\newcommand{\eal}{\end{align}}
\newcommand{\bi}{\begin{itemize}}
\newcommand{\ei}{\end{itemize}}
\newcommand{\ben}{\begin{enumerate}}
\newcommand{\een}{\end{enumerate}}
\newcommand{\bc}{\begin{center}}
\newcommand{\ec}{\end{center}}
\newcommand{\bt}{\begin{table}}
\newcommand{\et}{\end{table}}
\newcommand{\btb}{\begin{tabular}}
\newcommand{\etb}{\end{tabular}}

\newcommand{\bl}{\left}
\newcommand{\br}{\right}

\newcommand{\eV}{\mathrm{eV}}

\newcommand{\MeV}{\mathrm{MeV}}
\newcommand{\GeV}{\mathrm{GeV}}
\newcommand{\TeV}{\mathrm{TeV}}
\newcommand{\gr}{\mathrm{g}}

\newcommand{\cm}{\mathrm{cm}}

\newcommand{\yr}{\mathrm{yr}}

\newcommand{\abs}[1]{\ensuremath{\bl\vert{#1}\br\vert}}
\newcommand{\thavg}[1]{\ensuremath{\bl\langle {#1} \br\rangle}}
\newcommand{\anti}[1]{\ensuremath{\overline{#1}}}

\newcommand{\mO}{\mathcal{O}}

\newcommand{\mV}{\mathcal{V}}
\newcommand{\Ztwo}{\mathbb{Z}_2}
\newcommand{\sB}{\ensuremath{{\cancel{B}}}}

\newcommand{\sCP}{\ensuremath{{\cancel{CP}}}}
\newcommand{\eCP}{\ensuremath{\epsilon_{\rm CP}}}
\newcommand{\eCPp}{\ensuremath{\epsilon_{\rm CP}'}}

\newcommand{\chip}{\ensuremath{{\chi'}}}

\newcommand{\altp}[1]{\ensuremath{{#1^{(\prime)}}}}
\newcommand{\altpp}[2]{\ensuremath{{#1_{#2^{(\prime)}}}}}
\newcommand{\Lam}{\ensuremath{\Lambda}}

\newcommand{\LQCD}{\ensuremath{\Lambda_\mathrm{QCD}}}
\newcommand{\LQCDp}{\ensuremath{\Lambda_\mathrm{QCD}'}}

\newcommand{\veps}{\ensuremath{\varepsilon}}

\newcommand{\hc}{\ensuremath{\mathrm{h.c.}}}

\newcommand{\vs}{\ensuremath{\mathrm{vs}}}

\newcommand{\ds}{\ensuremath{\mathrm{ds}}}
\newcommand{\rh}{\ensuremath{\mathrm{rh}}}
\newcommand{\Pl}{\ensuremath{\mathrm{Pl}}}
\newcommand{\fo}{\ensuremath{\mathrm{fo}}}

\newcommand{\eq}{\ensuremath{\mathrm{eq}}}

\newcommand{\DNeff}{\ensuremath{\Delta N_\mathrm{eff}}}
\newcommand{\dm}{\ensuremath{\mathrm{dm}}}

\newcommand{\tot}{\ensuremath{\mathrm{tot}}}

\newcommand{\UV}{\ensuremath{\mathrm{UV}}}

\newcommand{\gsim}{\lower.7ex\hbox{$\;\stackrel{\textstyle>}{\sim}\;$}}
\newcommand{\lsim}{\lower.7ex\hbox{$\;\stackrel{\textstyle<}{\sim}\;$}}

\newcommand{\ignore}[1]{}

\begin{document}
\hspace{32em} {\small FERMILAB-PUB-24-0023-T-V}
\vspace{2em}

\title{A Closer Look in the Mirror: \\
Reflections on the Matter/Dark Matter Coincidence}

\author{Arushi Bodas}
\email{arushib@uchicago.edu}
\affiliation{Enrico Fermi Institute, University of Chicago, Chicago, IL 60637, USA}
\affiliation{Particle Theory Department, Fermilab, Batavia, Illinois 60510, USA}
\affiliation{Maryland Center for Fundamental Physics, Department of Physics, University of Maryland, College Park, MD 20742, USA}

\author{Manuel A. Buen-Abad}
\email{buenabad@umd.edu}
\affiliation{Maryland Center for Fundamental Physics, Department of Physics, University of Maryland, College Park, MD 20742, USA}
\affiliation{Department of Physics and Astronomy, Johns Hopkins University, Baltimore, MD 21218, USA}
\affiliation{Dual CP Institute of High Energy Physics, C.P. 28045, Colima, M\'{e}xico}

\author{Anson Hook}
\email{hook@umd.edu}
\affiliation{Maryland Center for Fundamental Physics, Department of Physics, University of Maryland, College Park, MD 20742, USA}

\author{Raman Sundrum}
\email{raman@umd.edu}
\affiliation{Maryland Center for Fundamental Physics, Department of Physics, University of Maryland, College Park, MD 20742, USA}

\begin{abstract}
    We argue that the striking similarity between the  cosmic abundances of baryons and dark matter, despite their very different astrophysical behavior, strongly motivates  the scenario in which dark matter resides within  a rich dark sector parallel in structure to that of the standard model. The near cosmic coincidence is then explained by  an approximate $\Ztwo$ exchange symmetry between the two sectors, where dark matter consists of stable dark neutrons, with matter and dark matter asymmetries arising via parallel WIMP baryogenesis mechanisms. Taking a top-down perspective, we point out that an adequate $\Ztwo$ symmetry necessitates solving the electroweak hierarchy problem in each sector, without our committing to a specific implementation. A higher-dimensional realization in the far UV is presented, in which the hierarchical couplings of the two sectors and the requisite $\Ztwo$-breaking structure arise naturally from extra-dimensional localization and gauge symmetries. We trace the cosmic history, paying attention to potential pitfalls not fully considered in previous literature. Residual $\Ztwo$-breaking can very plausibly give rise to the asymmetric reheating of the two sectors, needed to keep the cosmological abundance of relativistic dark particles below tight bounds. We show that, despite the need to keep inter-sector couplings highly suppressed after asymmetric reheating, there can naturally be order-one couplings mediated by TeV scale particles which can allow experimental probes of the dark sector at high energy colliders. Massive mediators can also induce dark matter direct detection signals, but likely at or below the neutrino floor.
\end{abstract}

\maketitle

\tableofcontents


\section{Introduction}
\label{sec:intro}

The identity of dark matter (DM) represents one of the greatest mysteries of Nature, whose resolution requires new physics beyond the Standard Model (BSM). Theories of DM range from those in which its physics is closely related to the Standard Model (SM), to those in which it is utterly alien. The most popular theories invoke an important but indirect DM-SM connection. For example, axion DM is tied to the axion solution to the Strong $CP$ Problem in which QCD generates the axion mass, while in the weak scale SUSY paradigm, Weakly Interacting Massive Particle (WIMP) DM is related to the SM by Supersymmetry (see, for example, Refs.~\cite{Hook:2018dlk,Fox:2019bgz} for recent reviews). And yet, such deep connections are not immediately evident in the observed properties of DM.

Here, we will take seriously the few intriguing hints we have of a more explicit connection based on the {\it observed structure} of DM and the SM:
\begin{enumerate}
  \item The observed cosmological abundance of dark and ordinary matter (or baryons, denoted by ``$b$'') are quite similar today, $\rho_\dm \approx 5 \rho_b$, suggesting an underlying connection in their origin. At least for the baryon density, its origin is quite subtle: in the hot early Universe it must involve almost equal numbers of baryons and antibaryons, with only a tiny ``matter-antimatter asymmetry'' (in terms of the ratio of its number density $n_b$ to the total entropy density $s$ in photons and neutrinos, $n_b/s \approx 0.14 \eta_b \approx 8.6 \times 10^{-11}$ \cite{Workman:2022ynf}, where $\eta_b \equiv n_b / n_\gamma$). While the symmetric component rapidly annihilates away, only the slight excess of baryons survives today as $\rho_b$.

  \item The real world is on a razor's edge in SM parameter space. Because the neutron-proton mass difference is so small compared to the electroweak scale, $m_n - m_p \approx 1~\MeV \ll v_{\rm weak}$, even a modest change in quark Yukawa couplings $\Delta y \sim y \ll 1$ could make the neutron lighter than the proton, stabilizing the neutron and making the proton short-lived. For example, if the proton was heavier than the neutron by $\gtrsim 3~\MeV$, the proton lifetime would be $< 1$ second, so protons would have disappeared before Big Bang Nucleosynthesis (BBN). Since in Nature there is no dineutron bound state \cite{Spyrou:2012zz}, baryonic matter would remain entirely in the form of free neutrons. These free neutrons would behave like an almost-ideal CDM.\footnote{Indeed, these neutrons only feel short-range nuclear forces and not long-range Coulomb interactions. Furthermore, as we discuss in detail in \Sec{subsec:nucl}, the lack of BBN means no atoms are ever formed. In particular they will satisfy standard CDM constraints, such as their self-interaction cross section and mass obeying $\sigma/m \lesssim 1~\cm/\gr$, from observations of the Bullet Cluster \cite{Markevitch:2003at,Randall:2008ppe,Robertson:2016xjh}.} Therefore, the abundance of neutron CDM in this hypothetical world would be essentially the same as the baryon abundance in our world, $\rho_{\rm neutron-CDM} \approx \rho_b$, given the same mechanism for the baryon-antibaryon asymmetry.
\end{enumerate}

Even in models in which the DM and baryon abundances arise through separately elegant but very different mechanisms, such as the misalignment mechanism for axion DM \cite{Abbott:1982af,Dine:1982ah,Preskill:1982cy,Hook:2018dlk} or leptogenesis for the baryon abundance \cite{Flanz:1994yx,Covi:1996wh,Covi:1996fm,Chen:2007fv,Davidson:2008bu,Bodeker:2020ghk}, the close similarity in their cosmic mass densities requires extreme and implausible fine-tuning in the {\it combined} parameter space. However, if DM belongs to a dark particle sector related to SM energy scales by new connecting symmetries and/or mandated by a solution to the electroweak hierarchy problem, the situation is somewhat improved. In the paradigm of Asymmetric Dark Matter (ADM) there is a conserved $U(1)$ global symmetry under which both the DM particles and our baryons and/or leptons are charged, which ensures similarity of the net DM and baryon number densities,  $n_\dm \sim n_b$ (the earliest versions of ADM appear in Refs.~\cite{Spergel:1984re,Nussinov:1985xr,Gelmini:1986zz,Barr:1990ca,Barr:1991qn,Kaplan:1991ah,Gudnason:2006ug,Gudnason:2006yj,Kaplan:2009ag}, for reviews see Refs.~\cite{Davoudiasl:2012uw,Petraki:2013wwa,Zurek:2013wia}). This does not fully explain why $\rho_\dm \approx 5 \rho_b$ because $\rho = m \, n$, so it also requires $m_\dm \approx 5 m_b \approx 5$ GeV. But if the dark sector occupies the same energy scales as the SM for symmetry or structural reasons, then such a DM mass is at least plausible. In the paradigm of WIMP DM, WIMP masses of the order the weak scale leads to the ``WIMP miracle'', in which thermal freezeout of WIMPs in the early Universe naturally leads to a DM abundance of the right ballpark, (see Refs.~\cite{Kolb:1990vq,Fox:2019bgz} for reviews)
\beq
    \rho_{\rm WIMP} \sim \rho_\dm \bl( \frac{10^{-9}~\GeV^{-2}}{\thavg{\sigma v}} \br) \ ,
\eeq
where $\thavg{\sigma v} \sim 10^{-9}~\GeV^{-2}$ is a typical order of magnitude for annihilation cross sections involving weak scale masses and couplings. But of course, this alone does not explain why $\rho_b$ should be similar to $\rho_{\rm WIMP}$. Furthermore, in the WIMP miracle, the DM abundance consists predominantly of a symmetric combination of WIMPs and their anti-particles (if these are even distinct), which is at odds with the ADM mechanism.

An attempt to relate baryogenesis to the DM WIMP miracle was developed in the framework of WIMP baryogenesis \cite{Cui:2012jh}. In addition to the stable DM WIMP, it requires an unstable but long-lived WIMP, the ``baryon parent'', which decays into baryons {\it after} its freezeout. As a result, the baryon abundance is given by
\beq
    \rho_b \sim \rho_\dm \, \eCP \frac{m_b}{m_\dm} \bl( \frac{10^{-9}~\GeV^{-2}}{\thavg{\sigma v}} \br) \ ,
\eeq
Here $\eCP$ is the $CP$-asymmetry generated by these decays, which can be as large as percent-level (with perturbative couplings). The observed baryon abundance can then be achieved by a somewhat smaller annihilation cross-section for the baryon parent than for the DM WIMP. 
A variant mechanism is to use long-lived WIMPs to achieve ADM \cite{Cui:2013bta}. While basing both DM and baryon abundances on WIMP-freezeout mitigates the fine-tuning involved in realizing their near coincidence, this tuning still persists. We can better understand this if we assume that some mechanism, such as ADM or an alternative, ensures comparable baryon and DM number densities, $n_b \sim n_\dm$. Then, the closeness in energy densities relies on the DM particle mass being close to the baryon mass. The composite baryon mass is dominated by its gluonic self-energy, which is set by the strong interaction RG-invariant scale of QCD, $m_b = {\cal O}(1) \times \LQCD \sim 1~\GeV$. This is parametrically very different from any (elementary) WIMP DM mass, so that a numerically similar mass would have to be a pure coincidence. Both this fact and hint (2) above inspire us to consider the possibility of a dark sector (DS) that is a close copy of the visible sector (VS), including the SM, but where the dark neutron is lighter than the dark proton and  makes up the observed DM of our Universe (such a possibility was first entertained, to the best of our knowledge, in Ref.~\cite{An:2009vq}). Throughout the rest of this paper, we denote with a prime ($\prime$) those fields and quantities associated with the dark sector. Note that the DS gauge forces only act on dark quarks and dark fermions, and the SM gauge forces only act on ordinary quarks and leptons. This leaves gravity as the dominant astrophysical interaction between DM and the SM, in agreement with observations.

Given that we are invoking a dark neutron dark matter mass of the order of $m_\dm = m_{n'} = {\cal O}(1) \times \LQCDp \sim \mO(\GeV)$, we should investigate how close the similarity between the two sectors needs to be. A central consideration is that $\LQCD$ is very sensitive to RG running over the vast hierarchy between the Planck scale and observable energies, so that having two such scales emerge close to each other in the IR is a strong requirement. To explore this, note that since the QCD theories are weakly coupled for most of the hierarchy we can use the one-loop approximation to the RG, to obtain
\beq
    \LQCD^{\frac{11 N_c}{3} - \frac{2 N_L}{3}} = M_\Pl^{ \frac{11 N_c}{3} - \frac{2 N_F}{3}} \, e^{-2 \pi/\alpha_s(M_\Pl)} \, \prod_{m_i > \LQCD} m_i^{2/3} \ . \label{eq:qcd}
\eeq
Here, $N_c$ denotes the number of colors of an $SU(N_c)$ gauge group for QCD, $N_F$ denotes the number of (effective) quark flavors significantly lighter than $M_\Pl$, and $N_L \leq N_F$ denotes the number of quark flavors lighter than $\LQCD$. We will study each type of input parameter individually, keeping other parameters identical between the two sectors, and check how close the selected parameter must be to keep $1 \lesssim \LQCDp/\LQCD \lesssim 10$.

We begin by assuming the $N_c = 3$, $N_F = 6$, and $N_L =3$ structure of SM QCD is shared by dark QCD, and that even the quark masses are equal in value. In this way, only $\alpha_s(M_\Pl)$ can differ between the two sectors. For the two QCD scales  to lie within an order of magnitude, the gauge couplings at the Planck scale have to be very close, $\Delta \alpha_s/\alpha_s \lesssim {\cal O}(10 \%)$, given that in the SM $\alpha_s(M_\Pl) \approx 0.02$. Famously, UV-equality of gauge couplings can arise in the context of unification and unified symmetries. Let us adopt such a scenario, with dark QCD unified with QCD somewhere near the Planck scale so that $\alpha_s'(M_\Pl) \approx \alpha_s(M_\Pl)$. We now focus on constraining $N_c'$ (keeping all other parameters identical). It is straightforward to check that the only way to then keep the QCD scales within an order of magnitude is to have $N_c' =3$, as in standard QCD. Similarly, focusing on $N_F'$ and $N_L'$ while fixing all other parameters, we find $N_F' = 6$ and $N_L' = 3$ as the only possibility, as in standard QCD. Finally, let us turn to the quark masses, keeping other parameters identical. In order to have $1 \lesssim \Lambda'/\Lambda \lesssim 10$, we must rescale the dark quarks to be heavier, by at most a factor $\sim {\cal O}(100)$ on (geometric) average. Given that the SM quark masses are set by diverse Yukawa couplings multiplied by the weak scale $v = 246~\GeV$, this strongly suggests that the dark quark masses have an analogous origin, with hierarchical Yukawa couplings to a dark Higgs sector with a comparable dark electroweak scale $v' \gtrsim v$. Note that $v'$ and the up- and down-quark Yukawas cannot be too large, otherwise the mass of the dark neutron DM will predominantly come from its constituent dark quarks instead of $\LQCDp$, which will spoil our explanation for the DM-baryon coincidence.

There is a very simple way to satisfy all these tight requirements, namely to have an approximate $\Ztwo$ symmetry exchanging the two sectors.\footnote{In this work we assume that the entire SM is replicated in the dark sector. In particular, there are dark electroweak interactions. As we shall see, these play a crucial role in ensuring the decay of undesirable heavy states such as the dark proton, and in ``eating'' the dark Higgs would-be Goldstone bosons, which would otherwise be massless and contribute to the number of extra relativistic degrees of freedom parametrized by $\DNeff$.} Such a symmetry has been considered in the past from different viewpoints, but we defer a comparison with the literature to the end of this introduction. In our context we can deduce that this $\Ztwo$ symmetry must not be exact because we need the dark light-quark Yukawa couplings to be shifted by $\gtrsim 10\%$ in order for the dark neutrons to be lighter than dark protons. However, it must be more closely respected by the gauge interactions, because we want $\Delta \alpha_s/\alpha_s \lesssim {\cal O}(10 \%)$ in the UV, as previously stated.

This raises a central question of the scale at which $\Ztwo$-breaking takes place: in the far UV (closer to the Planck scale) or in the IR (closer to TeV)? In the latter case, initiated by spontaneous breaking of $\Ztwo$, it is difficult with perturbative renormalizable dynamics to communicate {\it  hard} $\Ztwo$-breaking as large as $\sim {\cal O}(10 \%)$ to the gauge and Yukawa couplings, consistent with our other requirements.\footnote{Reference \cite{An:2009vq} is an interesting attempt in this direction of IR $\Ztwo$-breaking, but with somewhat different motivations and features.} We therefore choose to realize modest $\Ztwo$-breaking in the far UV, where non-renormalizable effective couplings may be UV-completed by dynamics beyond field theory associated with quantum gravity, such as string theory.

\begin{figure}[h!]
  \centering
  \includegraphics[width=0.7\linewidth]{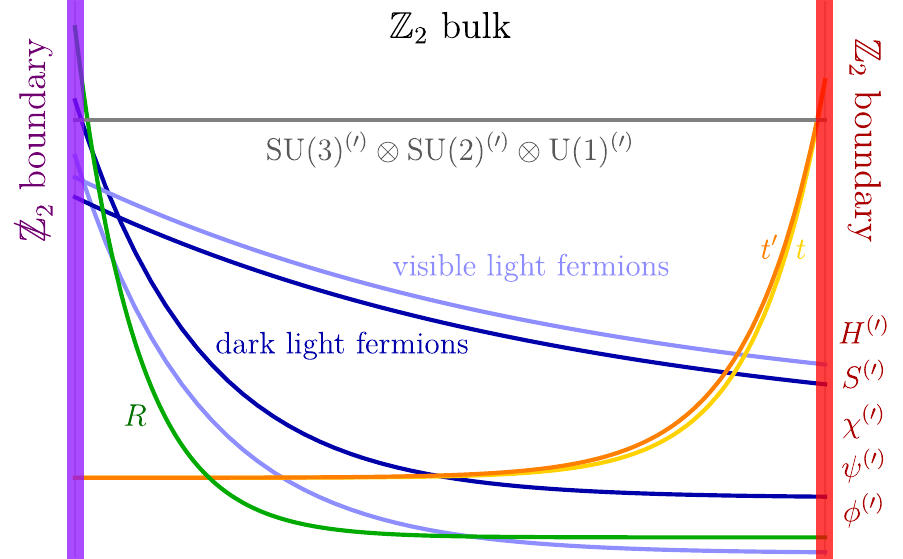}
  \caption{
  Sketch of the extra-dimensional realization of our model. The left-hand boundary of a compact extra dimension (in purple) violates with order one the $\Ztwo$ symmetry that exchanges visible and dark fields, while the right-hand boundary (in red) preserves it. In a manner analogous to Partial Compositeness, the zero-modes of the light fermions (blue curves) are exponentially peaked toward the left-hand boundary, while those of the top and its dark counterpart (yellow and orange curves) are peaked towards the right-hand boundary. This arrangement ensures $\Ztwo$-breaking ($\Ztwo$-symmetric) Yukawas for the light fermions (top-quarks). The gauge fields (gray line), live in the $\Ztwo$-symmetric bulk; any $\Ztwo$-violating differences between their visible and dark gauge couplings originate from the $\Ztwo$-breaking boundary and are volume-diluted, and therefore small. All the new fields involved in the WIMP baryogenesis mechanism and its dark version are localized on the right-hand boundary, thus ensuring $\Ztwo$-replicated couplings. The $\Ztwo$-singlet reheaton $R$ (green) is, however, peaked toward the left-hand boundary. Throughout our paper, we take the Kaluza-Klein scale to be close to the GUT scale $\sim 10^{15-16}$ GeV.
  }
  \label{fig:orbifold}
\end{figure}

Taking the $\Ztwo$ as a discrete unified (gauge) symmetry, the requisite structure of symmetry breaking can be simply achieved in an extra-dimensional framework, in the spirit of  orbifold-unification \cite{Kawamura:1999nj,Kawamura:2000ev,Hall:2002ea} (usually applied to grand-unified extensions of the SM gauge group). It simultaneously allows us to realize the attractive  Partial Compositeness mechanism\footnote{While this mechanism is usually realized in an extra-dimensional setting, as here, the  ``compositeness'' in its name arises from its CFT/AdS dual form (when the extra dimension is warped).} for generating hierarchical flavor structure \cite{Kaplan:1991dc,Grossman:1999ra,Gherghetta:2000qt,Huber:2000ie,Huber:2003tu}. For a brief account see Ref. \cite{Agashe:2005vg}. Our extra-dimensional set-up is illustrated in \Fig{fig:orbifold}. Two boundaries, one $\Ztwo$-breaking and the other $\Ztwo$-preserving (left and right, respectively), enclose a $\Ztwo$-symmetric extra-dimensional bulk. The VS and DS Higgses, as well as those fields required for WIMP (dark) baryogenesis and which will be introduced in detail in the next section, are localized on the $\Ztwo$-symmetric boundary. On the other hand, the SM chiral fermions, their dark counterparts, and a ``reheaton'', arise as zero-modes of the extra-dimensional compactification with a variety of extra-dimensional exponential wavefunctions. We take the visible and dark top-quarks to be peaked towards the $\Ztwo$-symmetric boundary, while the visible and dark light fermions, as well as the reheaton, are peaked towards the $\Ztwo$-breaking one. In this way, all the hierarchies in the couplings of these fields (\eg, those in the Yukawas) arise from the diverse wavefunction overlaps with the boundary-localized fields. Furthermore, this arrangement ensures that the $\Ztwo$ symmetry breaking predominantly impacts the light quarks and leptons and their Yukawa couplings at order one, while on the other hand leaving the top-quark Yukawa to be closely replicated in the dark sector. Finally, gauge field profiles are flat in the bulk (by higher-dimensional gauge invariance) making it straightforward to achieve the requisite $\Delta \alpha_s/\alpha_s \lesssim {\cal O}(10 \%)$. Our extra-dimensional structure, while important for enforcing the pattern of $\Ztwo$ symmetry and its breaking in the couplings, as well as their hierarchies, need only appear in the far UV, for example at a Kaluza-Klein scale typical of unification, $\Lambda_\UV \sim M_{\rm GUT} \sim 10^{15\text{--}16}~\GeV$. Given how high this scale is, the extra dimension does not play an explicit role in our model, but economically sets the expectations for our UV boundary conditions.

There is an important subtlety with this $\Ztwo$ symmetry due to the electroweak hierarchy problem. Famously, the SM requires extreme fine-tuning among couplings to arrange $v \ll M_\Pl$. If the $\Ztwo$ symmetry were {\it exact}, then the same fine-tuning would automatically take place in the dark sector, $v' = v \ll M_\Pl$. However, with even modest $\Ztwo$-breaking couplings in the UV, the dark sector would be so detuned that $v'$ would be orders of magnitude above $v$  without an independent extreme fine-tuning; we have argued against such large values of $v'$ above. We therefore conclude that for our scenario to be a plausible explanation of the DM-baryon coincidence, it {\it requires} that the hierarchy problem be solved in both sectors, by similar means. If the new physics solving the hierarchy problem in both sectors has masses $\sim m_{\rm NP}$, the most natural possibility is that $v \approx v' \sim m_{\rm NP}$. However, as is well known, in the most attractive models solving the hierarchy problem, new physics this light has been excluded experimentally. Evading a number of direct and indirect constraints typically requires $m_{\rm NP} \gtrsim 1\text{--}10~\TeV$, at the cost of modest tuning $\sim v^2/m_{\rm NP}^2$, the so-called ``little hierarchy problem''.  In this case, the $\Ztwo$-breaking detuned dark sector will most naturally have intermediate weak scale, $v < v' < m_{\rm NP}$. The new physics in each sector (at $\sim 1 \text{--} 10~\TeV$) will necessarily modify \Eq{eq:qcd} in detail, however the same qualitative conclusions are to be expected.

In this paper, we will study such a realization of $\Ztwo$ symmetry(-breaking) between a visible sector and a dark sector (VS and DS, respectively). We take the VS to consist of the SM as well as NP solving its hierarchy problem, and new particles that implement WIMP baryogenesis \cite{Cui:2012jh}, both at $\sim \mathcal{O}(1\text{--}10~\TeV)$.\footnote{The fields necessary to implement WIMP baryogenesis may be related to the NP solving the hierarchy problem. For example, in the context of a supersymmetric model the stop can act as the diquark $\phi$, which we introduce in the next section and is necessary for WIMP baryogenesis \cite{Cui:2012jh}; correspondingly, in the dark sector the dark stop plays the role of a dark diquark. Grand unification at a scale $M_{\rm GUT}$ can then proceed in each sector as it does in the MSSM.} The dark sector  consists of a copy of the SM gauge-Higgs-fermion structure and the dark equivalent of its hierarchy problem resolution, as well as dark particles implementing WIMP dark-baryogenesis. The resulting stable dark neutrons then constitute the dark matter of our Universe, with comparable but non-identical abundance to ordinary baryons due to the modest $\Ztwo$-breaking. Our analysis will not depend on the details of the solution to the hierarchy problem(s), but merely that such a solution is in place.

The cosmological history of our model will begin with initial conditions with both sectors in post-inflationary thermal equilibrium. The approximately $\Ztwo$-symmetric physics and initial conditions give rise to $\rho_\dm \sim {\cal O}(\rho_b)$. Thus, natural expectation given the symmetry is that the relativistic light species in both sectors (dark radiation, or DR) have comparable energy densities, translating to an effective number of ``extra neutrinos'' of $\DNeff \sim 7$ \cite{Kolb:1990vq}, in conflict with the current cosmological 95\% C. R. bound of $\DNeff \leq 0.284$ \cite{Planck:2018vyg}. However,  these densities can easily be modified by the decays of any long-lived massive particle connecting the two sectors. The decays of such a ``reheaton'' can be quite asymmetric if their decay kinematics depends on $v < v'$,  which are the scales most sensitive to $\Ztwo$-breaking (related to the hierarchy problem) \cite{Arkani-Hamed:2016rle,Chacko:2016hvu}. Such asymmetric reheating can elegantly reduce $\DNeff$ to within current bounds.

We have arrived at the paradigm of a $\Ztwo$-related dark sector with a parallel dark baryogenesis for dark matter, and the need to solve the two-sector hierarchy problem, primarily following the observations (1) and (2) discussed above. Similar scenarios have arisen in the literature, but tied to other non-DM motivations which tightly constrain their realizations as theories of DM and baryogenesis. Examples of these motivations include ``mirror" sectors in which the (dark) particle spectra are taken to be exactly $\Ztwo$-symmetric either as a matter of principle or in order to restore a fundamental parity symmetry in particle physics \cite{PhysRevD.47.456,Berezhiani:1995am,Mohapatra:2000rk,Berezhiani:2000gw,Foot:2003jt,Berezhiani:2005ek,An:2009vq,Lonsdale:2018xwd}, or Twin Higgs theories in which the dark sector mitigates the little hierarchy problem \cite{GarciaGarcia:2015fol,Craig:2015xla,GarciaGarcia:2015pnn,Terning:2019hgj,Farina:2015uea,Farina:2016ndq,Koren:2019iuv,Beauchesne:2020mih,Feng:2020urb,Bittar:2023kdl,Alonso-Alvarez:2023bat}. Even more broadly, models involving the unification of visible and dark gauge forces above some high energy scale \cite{Lonsdale:2014wwa,Lonsdale:2014yua,Ibe:2019ena,Murgui:2021eqf} or the NNaturalness \cite{Arkani-Hamed:2016rle,Easa:2022vcw} approach to the hierarchy problem, also give rise to interesting connections between dark and ordinary matter. Finally, there are other, non-$\Ztwo$--based approaches to the DM-baryon coincidence puzzle that rely on ideas such as the anthropic principle \cite{Linde:1987bx,Wilczek:2004cr,Hellerman:2005yi}, conformal field theories \cite{Bai:2013xga,Newstead:2014jva,Ritter:2022opo}, correlated production and masses \cite{Rosa:2022sym}, and the use of moduli to dynamically solve the problem \cite{Brzeminski:2023wza}. 

Our previous deliberations have brought us to a scenario more reminiscent of mirror dark sectors. However, exact mirror sectors are challenging in terms of breaking $\Ztwo$ at least at the level of cosmological initial conditions,\footnote{Although see, for example, Ref.~\cite{Cline:2021fdy}, where asymmetric reheating within an exact mirror model is achieved in the context of an inflationary model.} and in terms of satisfying astrophysical constraints on dark-atom DM and on dark light relics. Here, our sole motivation is to find a combined theory of dark matter- and matter-genesis which remains natural up to very high energies, untethered from other theoretical goals. Nevertheless, separate elements of our work have appeared in the literature. Dark baryon DM was first proposed in Ref.~\cite{PhysRevD.47.456} (dark neutron DM in Ref.~\cite{An:2009vq}, the necessity of avoiding dark BBN in Ref.~\cite{Higaki:2013vuv}), WIMP dark/visible baryogenesis was first considered in a Twin-Higgs context \cite{Farina:2016ndq}, while asymmetric reheating of the kind we use in this paper was first studied in Refs.~\cite{Arkani-Hamed:2016rle,Chacko:2016hvu}.

We can ask how such a $\Ztwo$ DM-baryogenesis mechanism can be experimentally tested. Of course, the requirement that the hierarchy problem is solved in each sector implies that new physics in each sector must appear at energies not too far above the weak scale---superpartners, for example. The baryon-parent WIMP of the SM sector could also be pair-produced at colliders, with spectacular long-lived (baryon-number violating) decays back to the SM \cite{Cui:2012jh,Cui:2014twa}. However, these unavoidable signals do not probe the dark sector. Minimally, as discussed above, the two sectors are only coupled through the reheaton, and that only very weakly. But we will also consider the most natural portals that might be present, such as a Higgs-dark Higgs coupling or a shared gauged $B-L$ which would allow us to probe the dark sector more readily. These portals will generically lead to dark particles production at high-energy colliders, as well as to potentially observable signals at DM direct detection experiments.

This paper is organized as follows. In Section ~\ref{sec:model} we introduce our minimal working model, building on the WIMP baryogenesis framework of Ref.~\cite{Cui:2012jh}, and describe its various parts. Section~\ref{sec:thermal} describes in detail the thermal history of our model and the various cosmological bounds on its parameters. We discuss the equations governing WIMP freezeout, asymmetric reheating, baryogenesis and confinement (in both sectors), the dangers of dark nucleosynthesis, and the abundance of cold dark relics other than the dark neutron dark matter. We devote Section~\ref{sec:pheno} to various phenomenological and observational consequences of our minimal model and two of its VS portal extensions (the Higgs-dark Higgs portal, and the massive $B-L$ gauge portal). Finally, we conclude in Section~\ref{sec:concl}.

\section{The Model}
\label{sec:model}

As stated in the introduction, our setting involves $\Ztwo$-related dark and visible sectors. In addition to the SM and its dark counterpart, each has the ingredients necessary for WIMP (dark) Baryogenesis to take place. A reheaton $R$ (a scalar singlet in our particular realization) bridges both sectors through its couplings to their Higgs fields \cite{Arkani-Hamed:2016rle,Chacko:2016hvu}, and is {\it not} duplicated.\footnote{For the most part, we will not need to explicitly detail the NP responsible for solving the hierarchy problem in each sector, but simply assume it is also present. Whenever we need to talk in more detail about this NP, we will be explicit about it and assume a token scenario, \eg~Supersymmetry.} In other words, the terms of the potential beyond the (dark) SM are those found in Ref.~\cite{Cui:2012jh}, plus the  reheaton-Higgs interactions:
\beq\label{eq:Vfull}
    \Delta \mV \equiv \mV_\vs + \mV_\ds + \mV_{\rh} \ ,
\eeq
where
\bea
    \mV_\vs & \equiv & \lambda_{ij} \phi d^{c\dagger}_i d^{c\dagger}_j + \veps_i \phi \chi u^c_i + \kappa_i \phi \psi u^c_i + \frac{1}{2} \alpha \chi^2 S + \beta_S S \abs{H}^2 + \frac{1}{2} m_\chi \chi^2 + \frac{1}{2} m_\psi \psi^2 + \frac{1}{2} m_S^2 S^2 + \hc \ , \nonumber\\
    \mV_\ds & \equiv & \lambda_{ij}' \phi' d^{c\prime\dagger}_i d^{c\prime\dagger}_j  + \veps_i' \phi' \chi' u^{c\prime}_i + \kappa_i' \phi' \psi' u^{c\prime}_i+ \frac{1}{2} \alpha' \chip^2 S' + \beta_S' S' \abs{H'}^2 + \frac{1}{2} m_{\chi'} \chip^2 + \frac{1}{2} m_{\psi'} \psi^{\prime 2} + \frac{1}{2} m_{S'}^2 S^{\prime 2} + \hc \ , \nonumber\\
    \mV_\rh & \equiv & \beta_R R \abs{H}^2 + \beta_R' R \abs{H'}^2
    + \frac{1}{2} m_R^2 R^2 \ . \label{eq:Veach}
\eea

\begin{figure}[h!]
  \centering
  \includegraphics[width=0.8\linewidth]{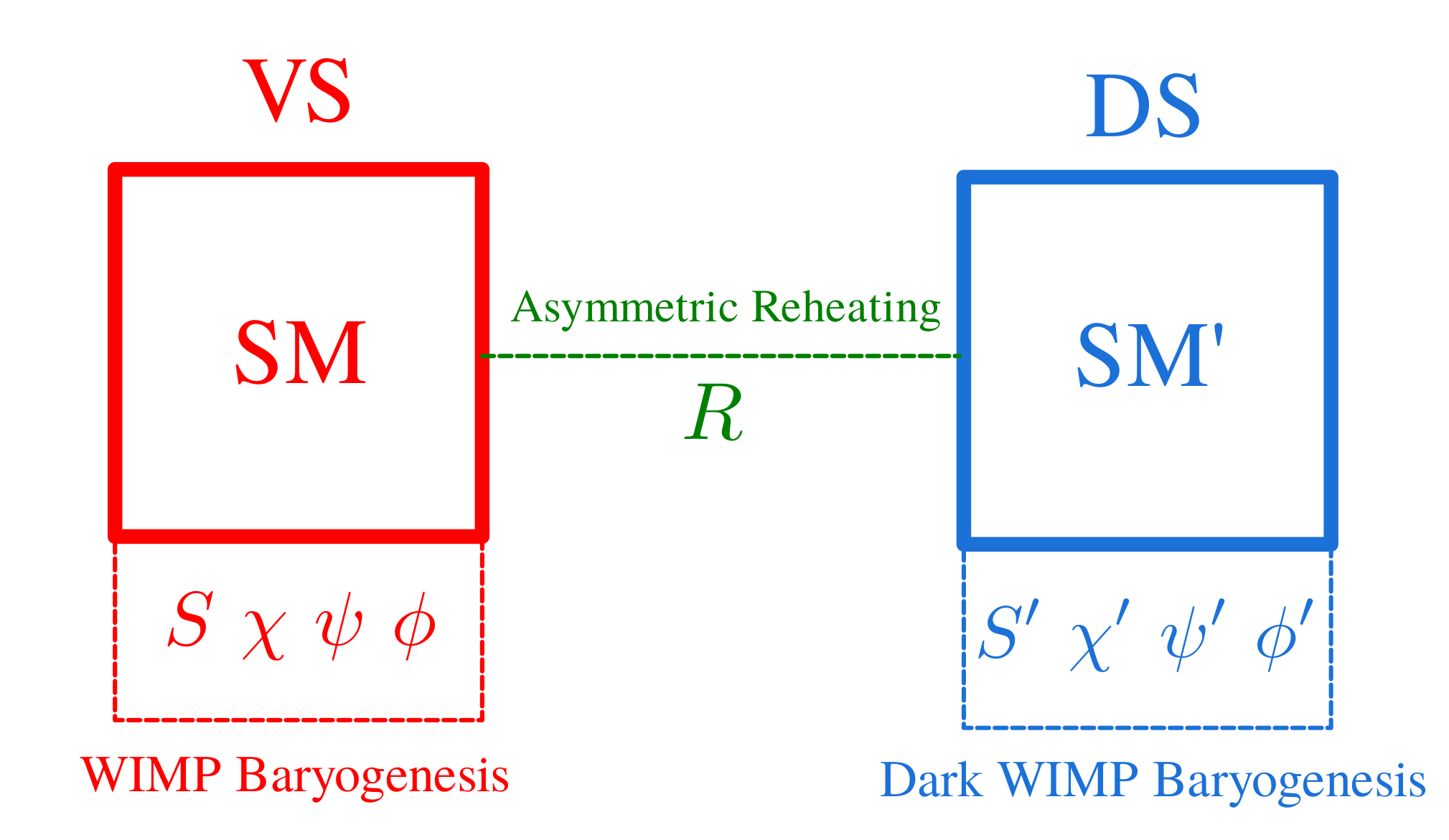}
  \caption{Sketch of the field content of our model. The visible and dark sectors are represented in red and blue, respectively. The VS (DS) contains the SM$^{(\prime)}$, as well as the $\altp{S}$, $\altp{\chi}$, $\altp{\psi}$, and $\altp{\phi}$ fields, necessary for (dark) WIMP Baryogenesis. The reheaton $R$, responsible for the asymmetrical reheating of the VS, bridges both sectors. With the exception of $\altp{\phi}$, which is a scalar diquark and is thus charged under the (dark) strong and hypercharge interactions, all these new fields are singlets under the SM$^{(\prime)}$ gauge forces.}
  \label{fig:fieldcontent}
\end{figure}

$\mV_\vs$ summarizes the VS WIMP baryogenesis mechanism of Ref.~\cite{Cui:2012jh}, where the $u^c$ and $d^c$ are the up- and down-type SM anti-quarks, $H$ is the SM Higgs, $\phi$ is a scalar diquark, $\psi$ and $\chi$ are Majorana fermions (WIMPs), with the latter being the long-lived baryon parent, and $S$ is the heavy mediator to the Higgs field ultimately responsible for the freezeout of $\chi$ relics.\footnote{We denote the baryon number-violating couplings of $\psi$ by $\altp{\kappa_i}$, instead of $\altp{y_i}$ previously used in Ref.~\cite{Cui:2012jh}, to distinguish them from the Yukawa couplings.} The indices $i$, $j$ run over the three fermion families. The DS counterpart of the WIMP baryogenesis in the VS is then found in $\mV_\ds$, with the dark equivalent of fields and quantities  denoted with primes ($\prime$) throughout the rest of this paper. $\mV_\rh$ includes the reheaton terms bridging the visible and dark sectors.\footnote{The $\altp{\beta_R} R \abs{\altp{H}}^2$ terms will induce a tadpole for the reheaton potential, as well as mass mixings between the Higgses and the reheaton. Given that the values for $\altp{\beta_R}$ we consider in this paper ($0.1~\eV$--$100~\eV$) are much smaller than all the other mass scales of our model, we will ignore the resulting VEV and treat mixing angles perturbatively.} The very small $\altp{\beta_R}$ terms control the asymmetric reheating favoring the VS (due to having $m_{h'} > m_h$) at late times.\footnote{We want to point out that several ``variants'' of this basic model could be viable as well, and can satisfy our demands for a robust explanation of the DM-baryon coincidence. For example, the $\altp{S}$ could be light and the $\altp{\chi}$ freezeout could instead take place via $\altp{\chi}\altp{\chi} \to \altp{S} \altp{S}$ annihilations; $\chi'$ and $\chi$ could be the same field, and so could $S'$ and $S$; or we could even have $S' = S = R$. The version we present in this paper, while not the most minimal in terms of the number of fields, is the most modular, with several of its parts identical to those in Ref.~\cite{Cui:2012jh} and replicated for the DS.} We summarize the non-SM field content in \Fig{fig:fieldcontent}, highlighting the role the new physics plays in WIMP Baryogenesis and asymmetric reheating.

While a detailed discussion of the model parameters will take place in the following sections, here we anticipate that $\altp{\veps}$ and $\altp{\beta_R}$ must be small. This is so that both the WIMP (dark) baryon parents $\altp{\chi}$ and the reheaton $R$ can be long-lived: decays out of equilibrium are necessary both for $\altp{\chi}$ to trigger baryogenesis, and for the visible and dark $R$ decay products to not thermally equilibrate with each other and thus avoid $\DNeff$ bounds. This can be achieved within the fundamental UV extra-dimensional framework given in the introduction, where all the new visible and dark fields needed for baryogenesis can be taken to be localized at the right-hand $\Ztwo$-symmetric boundary; see \Fig{fig:orbifold}. In this way they readily can have ${\cal O}(1)$ $\Ztwo$-symmetric couplings among themselves and to the Higgses and tops of the two sectors. The reheaton, however, is taken to be peaked towards the $\Ztwo$-breaking boundary. Since there is only one $\Ztwo$-singlet reheaton, its couplings $\beta_R$ and $\beta_R'$ to the $H$ and $H'$ Higgses will be identical and small, suppressed by the tiny overlap of the reheaton's bulk profile with the $\Ztwo$ boundary where the Higgses are located. The very small $\altp{\varepsilon}$ (coupling fields all localized on the same boundary) are technically natural and will therefore remain small under renormalization. We will show in \Sec{subsec:bl} how such tiny couplings can be fully natural from symmetry considerations when we extend our model with a gauged $B-L$ symmetry.

\begin{figure}[h!]
  \centering
  \includegraphics[width=0.99\linewidth,trim={1.7cm 2.7cm 1.7cm 3cm},clip]{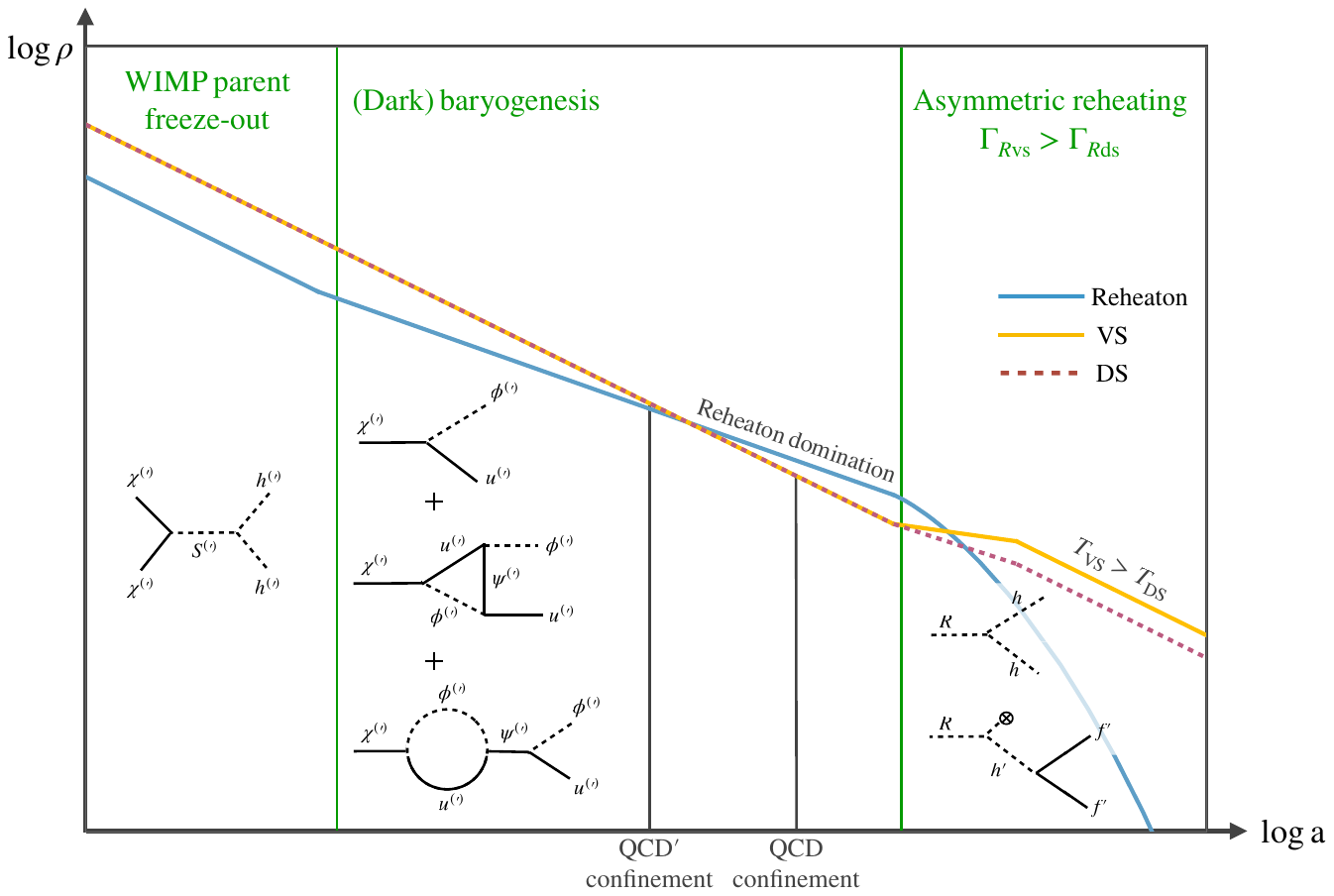}
  \caption{
  Sketch of the thermal history of our model, as a function of the scale factor $a(t)$. While the visible and dark sector densities (solid yellow and dashed purple lines, respectively) are the same, the (dark) baryon parents $\altp{\chi}$ undergo thermal freezeout via annihilations into (dark) Higgses (see diagram), resulting in identical abundances due to their $\Ztwo$-replicated couplings and masses. Some time after that, baryon-number washout processes shut off. Later, the relic $\altp{\chi}$ particles undergo $\sB$, $\sCP$ decays in the manner of WIMP baryogenesis through the Feynman diagrams shown. At some point around this time the era of reheaton domination begins, during which the reheaton $R$ preferentially decays into SM Higgses, due to the mass hierarchy $2 m_h < m_R < 2 m_{h'}$. Once this asymmetric reheating is completed, the VS dominates over the DS. Also around this time, (dark) confinement and hadron formation occur. The order in which (dark) baryogenesis, asymmetric reheating, and (dark) confinement take place is not important, as long as asymmetric reheating occurs after the onset of reheaton domination. After the VS has been reheated, and outside of the range of this figure, the remaining (dark) baryon-symmetric components annihilate or decay away, leaving only trace relics; and (dark) neutrino decoupling occurs, with BBN occurring shortly after. As we discuss in \Sec{subsec:nucl}, dark BBN does not take place.}
  \label{fig:history}
\end{figure}

A sketch of the thermal history that results from \Eqs{eq:Vfull}{eq:Veach} is shown in \Fig{fig:history}. We assume that at early times all particles are at high-temperature equilibrium through interactions mediated by additional, very heavy fields. As the Universe cools, the visible and dark sectors and the reheaton all decouple from each other, but with the same initial temperature. Having set this initial condition, these heavy fields are no longer relevant and play no further role in our story. As the Universe continues to expand, the (dark) baryon parents $\altp{\chi}$ undergo non-relativistic thermal freezeout through their annihilations into (dark) Higgses. As we will see explicitly in the next section, the approximate $\Ztwo$ symmetry relating the visible and dark sectors is enough to guarantee very similar $\altp{\chi}$ relic abundances. Some time after freezeout these parents decay in a $CP$- and baryon number-violating way ($\sCP$ and $\sB$, respectively) \`{a} la WIMP baryogenesis, via the first three terms of $\mV_\vs$ and $\mV_\ds$, also in an identical manner. Eventually the reheaton $R$ also decays through its Higgs couplings, preferentially reheating the VS since  $m_h < m_{h'}$. Hadrons are formed once the temperature drops below the confinement scale of the strong interactions, with the dark hadrons being $\mO(1)$ heavier than their visible partners due to $\LQCDp > \LQCD$. As discussed in the introduction, this mostly comes from detunings in the strong couplings $\alpha_s'$ and $\alpha_s$ at the fundamental UV scale. In addition, as was previosuly stated, we take $y_d' < y_u'$; the end result of the (dark) baryon asymmetry and hadron-antihadron annihilations is therefore the usual proton-dominated ordinary matter density, as well as dark-neutron dark matter. Finally, as we will show later, BBN occurs as usual in the VS, but fails to take place in the dark sector, leaving DM predominantly in the form of individual dark neutrons.

\section{Detailed Thermal History}
\label{sec:thermal}

In this section we describe in more detail the thermal history of our model, in chronological order. According to the rough picture we painted in the previous section, we detail the WIMP-like thermal freezeout of the (dark) visible baryon parent $\altp{\chi}$, the asymmetric reheating of the VS over the DS, (dark) baryogenesis, and (dark) confinement. We conclude with a short discussion on the annihilation of the remaining dark baryon-number symmetric relics and the absence of a dark BBN.

Anticipating the results of this section, we write the DM and baryon densities $\rho_\dm$ and $\rho_b$ as
\bea
    \rho_b & = & m_b n_b \simeq m_b \eCP Y_\chi \frac{s^\tot_0}{r_s} \ , \\
    \rho_\dm & = & m_\dm n_\dm \simeq m_{b'} \eCPp Y_\chip \frac{s^\tot_0}{r_s} \ ,
\eea
where $m_\dm = m_{b'} = m_{n'}$ is the mass of the dark neutron baryon, $Y_\altp{\chi}$ is the cold relic comoving number density resulting from the thermal freezeout of the $\altp{\chi}$ parents, $\altp{\eCP}$ the amount of $\sCP$ in the $\altp{\chi}$ baryogenesis decays, $s^\tot_0$ the total entropy density in the Universe today, and $r_s$ accounts for the out-of-equilibrium entropy dump due to the reheaton decays.

Since in our extra-dimensional setting the $\altp{\chi}$, $\altp{\psi}$, $\altp{\phi}$, $\altp{S}$, and $\altp{H}$ particles are localized on the $\Ztwo$-symmetric boundary, the couplings responsible for the $\altp{\chi}$ relic abundance and $\sCP$ decays are replicated in both the visible and dark sectors. As a result, $Y_\chi = Y_\chip$ and $\eCPp = \eCP$. The target values for the comoving number density and mass ratio can then be deduced from observations
\bea
    Y_\chi & \simeq & 8.6 \times 10^{-8} \ \bl( \frac{0.01}{\eCP} \br) \bl( \frac{r_s}{10} \br) \ , \label{eq:Ychi_bench} \\
    \text{and} \quad \frac{m_{b'}}{m_b}
    & \simeq & \frac{\rho_\dm}{\rho_b} \approx 5.3 \ , \label{eq:rho_ratio_bench}
\eea
where for the purposes of deriving these benchmarks we have taken the total entropy density to be equal to that in the visible sector, \ie~$s^\tot_0 \approx s_0 \approx 2.4 \times 10^{-38}~\GeV^3$; such an approximation is good to within $6\%$ for a $\DNeff = 0.3$ amount of DR. The plausibility of realizing these values within our scenario is explored in the subsequent sections.

\subsection{Freezeout and Asymmetric Reheating}\label{subsec:asymmReheat}

We will assume $m_\chi > m_h$, which means that the dominant channel for $\chi$ freezeout involves the heavy $S$-mediated $\chi\chi \to hh$ annihilations. Their cross section and resulting comoving number density are (see Ref.~\cite{Kolb:1990vq} for a review):
\bea
    \langle \sigma v \rangle_{\chi h} & \simeq & \frac{\alpha^2 \beta_S^2}{64 \pi m_S^4} \approx 10^{-15}~\GeV^{-2} \ \bl( \frac{\alpha}{0.5} \br)^2 \bl( \frac{\beta_S}{100~\GeV} \br)^2 \bl( \frac{10~\TeV}{m_S} \br)^4 \ , \\
    Y_\chi & \approx & 8 \times 10^{-8} \ \bl( \sqrt{\frac{200}{g_{*\chi\fo}^\tot}} \br) \bl( \frac{3~\TeV}{m_\chi} \br) \bl( \frac{10^{-15} ~ \GeV^{-2}}{\langle \sigma v \rangle_{\chi h}} \br) \bl( \frac{x_\fo}{11} \br) \ , \label{eq:comov_dens} \\
    \text{with} \quad x_\fo & \simeq & \ln\bl[ 0.2 \frac{g_\chi}{\sqrt{g_{*\chi\fo}^\tot}} m_\Pl m_\chi \langle \sigma v \rangle \br] - \frac{1}{2} \ln \bl[ \ln(...) \br] \ , \label{eq:xfo}
\eea
where $g_{*\chi\fo}^\tot$ is the total number of relativistic degrees of freedom at the freezeout temperature $T_{\chi\fo} = m_\chi/x_\fo$.\footnote{We denote by $g_{*X}$ the VS number of relativistic degrees of freedom in energy at some temperature $T_X$. $g_{*X}'$ does the same for the DS at $T_X'$, while $g_{*X}^\tot$ denotes the total degrees of freedom in both sectors (with respect to the VS temperature). Right before asymmetric reheating, we will often have $g_*' = g_*$ and $T' = T$; the only differences arise from the fact that the mass thresholds in the DS are typically larger than those in the VS. This means that the VS plays ``catch-up'' with the DS: $g_*'$ decreases with time before $g_*$ does, since the temperature $T'$ in the DS falls below some dark particle mass $m'$ before $T$ in the VS drops below $m$. For the most part, such subtleties play no role whatsoever in our scenario, and we ignore their mostly insignificant impact. Yet another simplification we make is that we will often not distinguish between the relativistic degrees of freedom in energy ($g_*$) and in entropy ($g_{*s}$), since they are the same throughout most of the history of the Universe. This is particularly true for the relatively high temperature of $T_{\chi\fo}$.} In \Eq{eq:xfo}, the ellipsis in the second term denotes the argument of the first. As long as $m_{\chi'} > m_{h'}$, dark parent $\chip\chip \to h' h'$ annihilations are analogous to $\chi \chi \to h h$, with the corresponding primed quantities.

\begin{figure}[h!]
  \centering
  \includegraphics[width=0.7\textwidth]{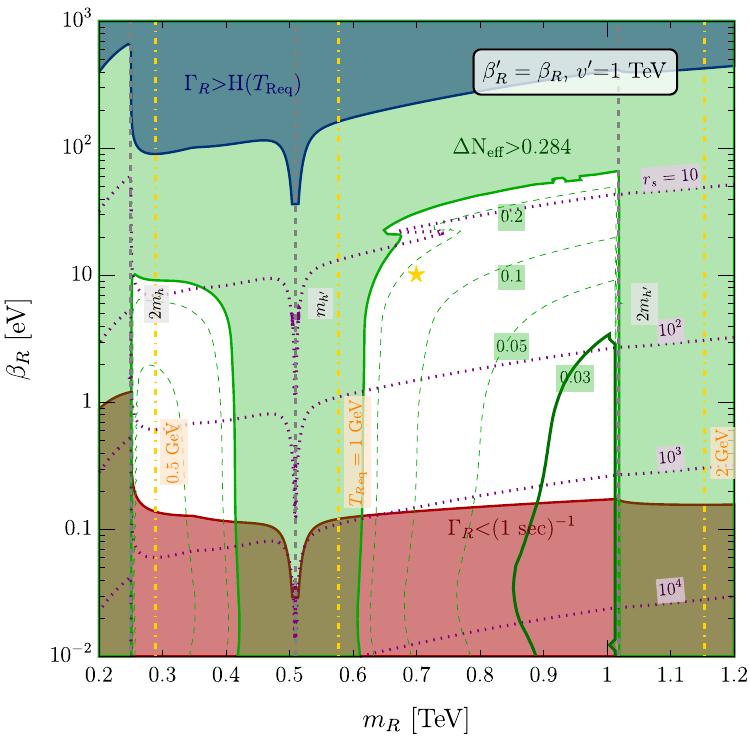}
  \caption{$m_R$--$\beta_R$ parameter space of the asymmetric reheating scenario. We have taken $\beta_R' = \beta_R$, $v' = 1~\TeV$ (\ie, $m_{h'} \approx 509~\GeV$), $y_b' = y_b$, and $g_{*sR\fo} = 2 \times 106.75$ as the total number of degrees of freedom at the time of reheaton relativistic freezeout (\ie, twice those of the SM). The blue (red) region corresponds to those points in parameter space where the reheaton decays before it dominates the energy density of the Universe (after BBN and neutrino decoupling). The green region corresponds to those points for which \Eq{eq:DNeff} violates the Planck constraints from TTTEEE+low-E+lensing+BAO for $\DNeff$ \cite{Planck:2018vyg}. The thin green contours correspond to several values of $\DNeff$ currently allowed by cosmological observations; the thick dark green contour ($\DNeff = 0.03$) roughly corresponds to the target sensitivities of future experiments \cite{Dvorkin:2022jyg}. The dotted purple lines represent different values of the entropy dump $r_s$ from \Eq{eq:rs}, while the dashed vertical orange lines indicate the temperature at which reheaton domination begins, according to \Eq{eq:TReq}. The dashed vertical grey lines represent the $m_R = 2 m_h$, $m_{h'}$, and $2 m_{h'}$ mass thresholds. The benchmark point used throughout this section, $m_R = 700~\GeV$ and $\beta_R = 10~\eV$, is represented by the yellow star.}
  \label{fig:arh}
\end{figure}

Having obtained the $\altp{\chi}$ relic abundances, let us move on to the reheaton decays, inspired by the work of Refs.~\cite{Arkani-Hamed:2016rle,Chacko:2016hvu}. These decays depend on $\beta_R$ and $\beta_R'$, the couplings of the reheaton $R$ to the visible and dark Higgses $H$ and $H'$, respectively ($\mV_\rh$ in \Eq{eq:Veach}). Since in our extra-dimensional setting $R$ is a $\Ztwo$-singlet bulk field, while $H$ and $H'$ live in the $\Ztwo$ boundary, we must have $\beta_R' = \beta_R$. Now well, for this late reheating to have a fighting chance at diluting away the DS energy density and thus allow our scenario to avoid the $\DNeff$ constraints, the reheaton must decay while it dominates the energy density of the Universe. For our weakly coupled reheaton (which freezes out at a temperature $T_{R\,\fo} \gg m_R$, while it is still relativistic), reheaton domination occurs after
$\rho_R = m_R n_R = (g_{*sR\eq}^\tot / g_{*sR\fo}^\tot) \zeta(3) m_R T_{R\eq}^3 / \pi^2 = \rho_{\rm rad} = g_{*R\eq}^\tot \pi^2 T_{R\eq}^4/30$,
\ie, for temperatures below
\beq
    T_{R\eq} \simeq 1~\GeV \ \bl( \frac{200}{g_{*sR\fo}^\tot} \br) \bl( \frac{m_R}{700~\GeV} \br) \ , \label{eq:TReq}
\eeq
In \Fig{fig:arh} we show the $m_R$--$\beta_R$ parameter space for our asymmetric reheating scenario, with $\beta_R' = \beta_R$ and $v' = 1~\TeV$. Several contours of $T_{R\eq}$, according to \Eq{eq:TReq}, are shown as dot-dashed vertical yellow lines.

As stated in the previous section, we need the reheaton decays to kinematically favor SM final states, thus causing the VS to be preferentially reheated and avoiding $\DNeff$ constraints. For such a kinematic suppression of the $R\to$ DS branching ratio to take place, we demand $2 m_{h'} > m_R > 2 m_h$. Within this range the dominant decay channel is $R \to hh$, involving SM Higgses, while $R \to b' \anti{b'}$ is the largest DS channel, as $R$ decays into two heavier dark Higgses is forbidden. The sum of these visible and dark sector contributions gives the total reheaton decay rate $\Gamma_R = \Gamma_{R\vs} + \Gamma_{R\ds}$, with contributions mostly from $R \to hh$, $R \to t \anti{t}$ and $R \to b \anti{b}$, and from $R \to b' \anti{b'}$, respectively:
\beq
    \Gamma_{R\vs} \simeq \Gamma_{Rh} + \Gamma_{Rt} + \Gamma_{Rb} \ , \quad \text{and} \quad \Gamma_{R\ds} \simeq \Gamma_{Rb'} \ , \\ \label{eq:Gamma_vsds}
\eeq
\bea
    \text{with} \quad \Gamma_{Rh} & = & \frac{\beta_R^2}{16 \pi m_R} \, \sqrt{1 - \frac{4 m_h^2}{m_R^2}} \ , \label{eq:Gamma_Rh} \\
    \text{and} \quad \Gamma_{R\altp{f}} & = & \frac{3}{16 \pi} \frac{m_R 
 \bl( \altp{\beta_R} \altp{v} \altp{y_f} \br)^2 }{\bl( m_R^2 - \altpp{m^2}{h} \br)^2 + \altpp{m^2}{h} \altpp{\Gamma^2}{h}} \bl( 1 - \frac{4 \altpp{m^2}{f}}{m_R^2} \br)^{3/2} \ , \label{eq:Gamma_Rf}
\eea
where $\altp{f}$ denotes the final state fermions of the VS and DS channels respectively, namely $t$, $b$, and $b'$; and $\altpp{\Gamma}{h}$ denotes the width of the (dark) Higgs. The $\Ztwo$ symmetry implies  $m_{h'} = (v' / v) \, m_h$; in a similar vein we take $\Gamma_{h'} = (v' / v) \, \Gamma_h$. Assuming $m_R, m_{h'} \gg v, m_h, \altpp{m}{f} \gg \altpp{\Gamma}{h}$, we can simplify these expressions further:
\bea
    \Gamma_{R\vs} & \approx & \Gamma_{Rh} \approx \frac{\beta_R^2}{16 \pi m_R} \approx 3 \times 10^{-21}~\GeV \ \bl( \frac{\beta_R}{10~\eV} \br)^2 \bl( \frac{700~\GeV}{m_R} \br) \ , \quad \text{and} \label{eq:Gamma_Rh_appx} \\
    \Gamma_{R\ds} & \simeq & \Gamma_{Rb'} \approx \frac{3 m_R \beta_R^{\prime 2} v^{\prime 2}y_b^{\prime 2}}{16 \pi \bl( m_R^2 - m_{h'}^2 \br)^2} \approx 5 \times 10^{-23}~\GeV \, \bl( \frac{\beta_R'}{10~\eV} \br)^2 \bl( \frac{m_R}{700~\GeV} \br) \bl( \frac{y_b'}{y_b} \br)^2 \bl( \frac{1~\TeV}{v'} \br)^2 \frac{0.8}{\bl( m_R^2/m_{h'}^2 - 1 \br)^2} \label{eq:Gamma_Rf_appx} \ .
\eea

Note that if $m_R = m_{h'}$ the reheaton decay rate into the DS (via final state dark bottom-quarks) is resonantly enhanced, which is exactly the opposite of what we hope to achieve with our asymmetric reheating mechanism. We thus anticipate that the parameter space around such a resonance will be severely constrained. Indeed, we can use \Eqs{eq:Gamma_Rh_appx}{eq:Gamma_Rf_appx} to determine how far away from the resonance the reheaton mass $m_R$ should be in order to have $\Gamma_{R\vs} \gg \Gamma_{R\ds}$ (and thus $\Gamma_R \approx \Gamma_{R\vs} \approx \Gamma_{Rh}$):
\beq
    \abs{m_R - m_{h'}} \gg 20~\GeV \ \bl( \frac{\beta_R'}{\beta_R} \br) \bl( \frac{v'}{1~\TeV} \br) \bl( \frac{y_b'}{y_b} \br) \ .
\eeq

In addition to decaying only after it dominates the energy density of the Universe, the reheaton must also decay before neutrino decoupling and BBN (when the Universe was roughly $1$ second old), since otherwise both processes will be significantly modified by the presence of extra DS radiation. More concretely, we require $(1~\sec)^{-1} < \Gamma_R < H(T_{R\eq})$; the red region of parameter space in \Fig{fig:arh} violates the first inequality, while the blue region corresponds to those points that violate the second. The mass thresholds $2 m_h$, $m_{h'}$, and $2 m_{h'}$, where $\Gamma_R$ changes most dramatically, are shown as dashed vertical grey lines in \Fig{fig:arh}; the impact that the $m_R = m_{h'}$ resonance has on these bounds is of special note. Since $\beta_R' = \beta_R$, these constraints on $\Gamma_R$ can be turned into simple analytic bounds on $\beta_R$ in the $\Gamma_{R\vs} \gg \Gamma_{R\ds}$ limit:
\beq
    0.1~\eV \ \bl( \frac{m_R}{700~\GeV} \br)^{1/2} < \beta_R < 300~\eV \ \bl( \frac{200}{g_{*sR\fo}^\tot} \br)^{3/4} \bl( \frac{m_R}{700~\GeV} \br)^{3/2} \ , \label{eq:arh_cond}
\eeq
having taken $g_{*R\eq}^\tot \lesssim g_{*sR\fo}^\tot \approx 200$, and away from any mass thresholds.

Asymmetric reheating results in different energy densities for the visible and dark sectors. These are simple to derive in the instantaneous decay approximation, where reheaton decays are assumed to be negligible up until the time $\tau$ at which $H(\tau) = \Gamma_R$; at that point the reheaton instantly reheats both sectors. Generalizing the results of Refs.~\cite{Kolb:1990vq} and \cite{Chacko:2016hvu}, the energy density $\altp{\rho_\rh}$ in the VS (DS) at the time of reheating is given by
\beq
    \altp{\rho_\rh} \simeq \rho_{R\tau} \bl[ \altp{\mathrm{BR}} + \frac{\pi^2/30}{\bl( \zeta(3)/\pi^2 \br)^{4/3}} \bl( \frac{g_{*sR\fo}^{(\prime)4}}{\altp{g_{*\tau}}} \br)^{1/3} \bl( \frac{\rho_{R\tau}}{m_R^4} \br)^{1/3} \br] \ , \label{eq:rho_rh}
\eeq
where
$
    \rho_{R\tau} \equiv 3 m_\Pl^2 \Gamma_R^2
$
is the reheaton energy density at $\tau$, $\altp{g_{*\tau}} \approx \altp{g_{*s\tau}}$ are the relativistic degrees of freedom of the VS (DS) at $\tau$ (with $\altp{g_{*\tau}} \approx 12$ being a typical value for our benchmarks), and $\altp{\mathrm{BR}}$ is the reheaton branching ratio to the VS (DS). The two terms of \Eq{eq:rho_rh} account for the two contributions to $\altp{\rho_\rh}$: the first comes from the decays of the reheaton $R$, and the second from the primordial radiation, redshifted from the time of $R$-freezeout (since $T_\tau^4 \sim T_{R\fo}^4 a_{R\fo}^4/a_\tau^4 \sim n_{R\tau}^{4/3} \sim \rho_{R\tau} \, (\rho_{R\tau}/m_R^4)^{1/3}$) \cite{Chacko:2016hvu}. From \Eq{eq:rho_rh} one can of course find the reheating temperature $\altp{T_\rh}$ from $\pi^2 \altp{g_{*\rh}} T_\rh^{(\prime)4} / 30 = \altp{\rho_\rh}$.

From the ratio $\rho_\rh' / \rho_\rh$ one can finally derive how much DR is there in the DS, parameterized as the equivalent number $\DNeff$ of extra neutrino families:
\bea
    \DNeff \simeq 7.4 \times \mathcal{G} \ \frac{\rho_\rh'}{\rho_\rh} \ , \quad \text{where} \quad \mathcal{G} \equiv \bl( \frac{g_{*0}'}{g_{*0}} \frac{g_{*\rh}}{g_{*\rh}'} \br) \bl( \frac{g_{*s0}}{g_{*s0}'} \frac{g_{*s\rh}'}{g_{*s\rh}} \br)^{4/3} \approx 1 \ . \label{eq:DNeff}
\eea
Contours of $\DNeff = 0.2, \, 0.1, \, 0.05$ are also shown as thin dashed green lines in \Fig{fig:arh}, while the 95\% C. R. bound on DR coming from the Planck satellite, $\DNeff \leq 0.284$ \cite{Planck:2018vyg}, corresponds to the shaded green region. Note that below $m_R = 2 m_h$ the $R \to hh$ VS channel is closed, and the reheaton visible branching ratio involves only fermions, severely limiting how much the VS is favored over the DS during reheating; hence the Planck constraint is quickly violated below this threshold. On the other hand, above $m_R = 2 m_{h'}$ the DS channel $R \to h' h'$ is open, which very quickly becomes an efficient way for the DS to be reheated. Furthermore, note how the region around $m_R = m_{h'}$ is ruled out, due to the resonant reheaton decays into dark bottom-quarks, which give a forbiddingly large DS branching ratio (see \Eq{eq:Gamma_Rf}). Overall, we see that $\mO(1)$ of the $2 m_h \leq m_R \leq 2 m_{h'}$ interval of interest is ruled out by the Planck bound on $\DNeff$, while $\mO(1)$ still remains open; close to two orders of magnitude in $\beta_R$ are also free from any constraint. Finally, we would like to remind the reader that the previous discussion on $\DNeff$ is moot if the VS and DS are still thermally coupled by the time the reheaton decays, since they would quickly equilibrate and share the same temperature. To be safe, we {\it conservatively} demand for both sectors to decouple before reheaton domination.\footnote{Of course, a more relaxed condition is to simply demand decoupling before the reheaton decays.} Since in our minimal model only the reheaton couples the two sectors, and that very weakly, this condition is easily satisfied in the parameter space of interest.

In the $\Gamma_{R\vs} \gg \Gamma_{R\ds}$ limit $\mathrm{BR} \approx 1$ and we can ignore the much smaller second term in \Eq{eq:rho_rh} for $\rho_\rh$; we must keep both terms for $\rho_\rh'$. In this limit the VS reheating temperature $T_\rh$ and the amount of DR $\DNeff$ are given by
\beq
    T_\rh \simeq \bl( \frac{90 \Gamma_R^2 m_\Pl^2}{g_{*\rh} \pi^2} \br)^{1/4} \approx 70~\MeV \ \bl( \frac{16}{g_{*\rh}} \br)^{1/4} \bl( \frac{\beta_R}{10~\eV} \br) \bl( \frac{700~\GeV}{m_R} \br)^{1/2} \ , \label{eq:Trh_appx}
\eeq
\bea
    \DNeff & \simeq & 0.1 \ \bl( \frac{\mathrm{BR}'}{0.02} \br) + 0.07 \ \bl( \frac{g_{*R\fo}'}{100} \br)^{4/3} \bl( \frac{12}{g_{*\tau}} \br)^{1/3} \bl( \frac{\beta_R}{10~\eV} \br)^{4/3} \bl( \frac{700~\GeV}{m_R} \br)^2 \ , \label{eq:DNeff_appx} \\
    \text{with} \quad \mathrm{BR}' & \simeq & \frac{\Gamma_{Rb'}}{\Gamma_{Rh}} \approx \ 0.02 \ \bl( \frac{\beta_R'}{\beta_R} \br)^2 \bl( \frac{m_R}{700~\GeV} \br)^2 \bl( \frac{y_b'}{y_b} \br)^2 \bl( \frac{1~\TeV}{v'} \br)^2 \frac{0.8}{\bl( m_R^2/m_{h'}^2 - 1 \br)^2} \ . \label{eq:BRp_appx}
\eea

Finally, we turn our attention to the entropy dump that results from the asymmetric reheating. The ratio $r_s$ of final-to-initial entropy per comoving volume, generated by the out-of-equilibrium reheaton decays during reheaton domination, is given by (see Ref.~\cite{Kolb:1990vq} for a review)
\bea
    r_s & \simeq & \bl[1 + 1.45 \ \frac{\bl( \zeta(3) / \pi^2 \br)^{4/3}}{2 \pi^2 / 45} \bl( \frac{g_{*s\rh}^\tot}{\bl(g_{*sR\fo}^\tot\br)^4} \br)^{1/3} \bl( \frac{m_R^4}{\rho_{R\tau}} \br)^{1/3} \br]^{3/4} \nonumber\\
    & \approx & 20 \ \bl( \frac{g_{*s\rh}^\tot}{17} \br)^{1/4} \bl( \frac{200}{g_{*sR\fo}^\tot} \br) \bl( \frac{10~\eV}{\beta_R} \br) \bl( \frac{m_R}{700~\GeV} \br)^{3/2} \ , \label{eq:rs}
\eea
where the last equality was obtained in the $\Gamma_{R\vs} \gg \Gamma_{R\ds}$ limit, and by dropping the first term inside the square brackets in the first line. In \Fig{fig:arh} we show the $r_s$ contours as a function of $m_R$ and $\beta_R$ as dotted purple lines.

Therefore, as shown by \Eqs{eq:comov_dens}{eq:rs}, our model readily accommodates the realistic  benchmark values of $Y_\chi$ and $r_s$ in \Eq{eq:Ychi_bench}. Furthermore, \Eqs{eq:DNeff}{eq:DNeff_appx} imply that our model can also satisfy the $\DNeff$ constraints through a modest asymmetric reheating and yet leave a potentially observable signal in future CMB and large-scale structure experiments \cite{Dvorkin:2022jyg} such as the Simons Observatory \cite{SimonsObservatory:2018koc}, CMB-S4 \cite{Abazajian:2019eic}, CMB-HD \cite{Nguyen:2017zqu,Sehgal:2019nmk,Sehgal:2019ewc,Sehgal:2020yja}, MegaMapper \cite{Schlegel:2019eqc}, and PUMA \cite{PUMA:2019jwd}. The combined future sensitivity of these experiments, $\DNeff \simeq 0.03$, is shown as the solid dark green contours in \Fig{fig:arh}. The conditions for our asymmetric reheating scenario to work are summarized in \Fig{fig:arh}, which plots the $m_R$--$\beta_R$ parameter space for $v' = 1~\TeV$ and $\beta_R' = \beta_R$.

\subsection{(Dark) Baryogenesis and Washout}
\label{subsec:baryo}

As in Ref.~\cite{Cui:2012jh}, the $\sCP$ and $\sB$ out-of-equilibrium decays of the $\chi$ baryon parents are responsible for baryogenesis. The decay rate and the amount of $CP$-asymmetry generated are \cite{Covi:1996wh,Chen:2007fv,Davidson:2008bu,Feng:2013zda}\footnote{The calculation proceeds as in the non-SUSY case studied in Ref.~\cite{Covi:1996wh}, with the vertex and self-energy loops contributing as $\mathcal{F}_V(m_\psi^2 / m_\chi^2) + 3 \mathcal{F}_S(m_\psi^2 / m_\chi^2)$, where $\mathcal{F}_V(x) = \sqrt{x}\bl[ 1 - (1+x) \ln (1 + x^{-1}) \br]$, and $\mathcal{F}_S(x) = \frac{1}{2} \sqrt{x} (1-x)^{-1}$ (and we have taken $x \gg 1$). Note that we have included a factor of $3$ due to the colors in the self-energy loop, and $\mathcal{F}_S(x)$ here differs from that in Ref.~\cite{Covi:1996wh} by a factor of $1/2$, accounting for the fact that the fields running in our loops are not $SU(2)$ doublets. See also Ref.~\cite{Feng:2013zda}.}
\bea
    \Gamma_\chi & \simeq & \frac{m_\chi}{8 \pi} \sum\limits_i \abs{\veps_i}^2 \sim 3 \times 10^{-19}~\GeV \bl( \frac{\veps}{5 \times 10^{-11}} \br)^2 \bl( \frac{m_\chi}{3~\TeV} \br) \ , \label{eq:Gamma_chi} \\
    \text{and} \quad \eCP & \simeq & \frac{2}{8 \pi} \, \frac{ \mathrm{Im} \bl\{ \bl( \sum\limits_i \veps_i \kappa_i^* \br)^2 \br\} }{\sum\limits_i \abs{\veps_i}^2} \, \frac{m_\chi}{m_\psi} \sim 0.02 ~ \bl( \frac{\kappa}{1} \br)^2 \bl( \frac{m_\chi}{3~\TeV} \br)  \bl( \frac{10~\TeV}{m_\psi} \br) \ ; \label{eq:eCP}
\eea
with $\veps \ll 1$, and $m_\psi > m_\chi > m_\phi$. The same is true, {\it mutatis mutandis}, for $\chip$. We would like to point out that, in order for the visible and dark WIMP baryogenesis mechanisms to be $\Ztwo$-symmetric (\ie, $\eCPp = \eCP$), we need $\kappa_i' = \kappa_i$ in the $\altp{\kappa_i} \altp{\phi} \altp{\psi} {u_i^c}^{(\prime)}$ terms of \Eq{eq:Veach}. In our extra-dimensional setting, illustrated in \Fig{fig:orbifold}, we expect the $\altp{\kappa_{1,2}}$ coefficients to be suppressed by the exponential tail of the light fermion zero-modes, while the $\Ztwo$-symmetric third-generation couplings $\kappa_3' = \kappa_3$ will remain large. Likewise, the same extra-dimensional geography guarantees $\veps_3' = \veps_3 \gg \altp{\veps_{1,2}}$, which means that $\sum\limits_i \altp{\veps_i} \kappa_i^{*(\prime)} \approx \altp{\veps_3} \kappa_3^{*(\prime)}$ and thus, according to \Eq{eq:eCP}, $\eCPp \approx \eCP$.

In order for baryogenesis to take place $\chi$ must decay out of equilibrium (\ie, $\Gamma_\chi < H(T_{\chi\fo} \sim m_\chi/11)$ for decays after freezeout). Furthermore, these decays must occur before BBN and neutrino decoupling (\ie, $\Gamma_\chi > (1~\sec)^{-1}$), in order to avoid spoiling these processes. These constraints mean that
\beq
    7 \times 10^{-14} \bl( \frac{3~\TeV}{m_\chi} \br)^{1/2} < \veps < 3 \times 10^{-8} \bl( \frac{g_{*\chi\fo}}{200} \br)^{1/4} \bl( \frac{m_\chi}{3~\TeV} \br)^{1/2} \ ;
    \label{eq:epsilon_bounds}
\eeq
the same is true for their dark counterparts.

In addition to these requirements, it is imperative that our hard-earned baryon number asymmetry not be washed out by various $\sB$ processes. Below we briefly discuss the conditions under which this is the case, which have been previously considered in Ref.~\cite{Cui:2012jh}:

\paragraph*{High-temperature ($T > \Lam$) washout:} The typical temperature below which the baryon number washout processes shut off is $T_w \lesssim T_{\fo} \sim m/\mO(10) \sim \mO(100~\GeV)$, where $m$ is $m_\phi$ or $m_\psi$. This is because of the Boltzmann exponential suppressions involved in these processes (as is the case for inverse decays), or because of the high-power of the temperature on which the rate in question depends ($T^{11}$ for $3 \to 3$ scattering, for example), or simply because this is the scale below which sphaleron processes become insignificant.

\paragraph*{Low-temperature ($T < \Lam$) washout:} $\Delta B = 2$ baryon-antibaryon oscillations could wash out baryon number at low temperatures. Reference~\cite{Cui:2012jh} considered several bounds stemming from the time of the QCD phase transition or of BBN, which are however straightforward to satisfy. In fact, tighter constraints on $\Delta B = 2$ processes arise from present-day oscillation and dinucleon-decay experiments. Improvements in these  present a potential avenue for future discovery, which we briefly discuss in 
\Sec{subsec:vs_pheno}.



Finally, we would like to say a few words about the chronology of our thermal history. It turns out that it is actually unimportant whether the asymmetric reheating occurs before or after the $\altp{\chi}$ decays. Indeed, all that matters is that both take place sufficiently late, after the onset of reheaton-domination, after $\altp{\chi}$ freezeout, and after baryon number washout processes cease to be important. For the example benchmark parameters we have chosen, the reheaton decays only after it dominates the energy density of the Universe, and the thermal bath has reached a temperature $T_{R\eq} \approx 1~\GeV$ (see \Eq{eq:TReq}); on the other hand, \Eq{eq:Gamma_chi} ensures $\altp{\chi}$ decays shortly after that.

\subsection{Confinement and the (Dark) Baryon Mass}
\label{subsec:confine}

In our scenario, the explanation of the DM-baryon coincidence relies on most of the dark baryon mass coming from the dark confinement scale, just like in the SM. In other words, we need $m_{b'} / m_b \simeq \LQCDp / \LQCD$. As stated in the introduction, both the visible and dark sectors must have $N_c = 3$ colors, $N_F = 6$ quark flavors, and $N_L = 3$ light quarks, which means that the 1-loop RG equations relating $\LQCD$ to an arbitrary $\Lambda_\UV$ scale are
\bea
    \Lambda_\UV^{7+B_{\rm NP}} & = & A \, \LQCD^9 \, \bl( m_c m_b m_t \br)^{-2/3}
    \!\!\!\!\!\!\!\!
    \prod\limits_{m_t < m_i < \Lambda_\UV}
    \!\!\!\!\!\!\!\!
    m_i^{b_i} \ , \quad \text{where} \\
    A & \equiv & \exp\bl[ \frac{2 \pi}{\alpha_{s,\UV}}\br] \ , \quad \text{and} \quad B_{\rm NP} =
    \!\!\!\!\!\!\!\!
    \sum\limits_{m_t < m_i < \Lambda_\UV}
    \!\!\!\!\!\!\!\!
    b_i \ ,
\eea
with the $i$ index running over any new colored particles heavier than the top quark, with a $b_i$ contribution to the 1-loop RG running of $\alpha_s$. Analogous expressions for the DS scale $\LQCDp$ and its associated dark colored states are easily obtained by putting primes in the corresponding places.

Assuming that any other color-charged states besides the visible and dark quarks have $\Ztwo$-symmetric masses between both sectors, we can write the ratio of confinement scales as
\bea
    \frac{\LQCDp}{\LQCD} & \simeq & \frac{A}{A'} \bl( \frac{m_{c'} m_{b'} m_{t'}}{m_c m_b m_t} \br)^{2/27} = \exp\bl[ \frac{2\pi}{9} \frac{\delta_{\alpha_s}}{\alpha_{s,\UV}} \br] \bl( \bl( \frac{v'}{v} \br)^3 \ \frac{y_c' y_b' y_t'}{y_c y_b y_t} \br)^{2/27} \ , \quad \text{or} \label{eq:lambda_ratio1} \\
    \frac{\LQCDp}{\LQCD} & \simeq & \bl( \frac{\Lambda_\UV}{\LQCD} \br)^{\delta_{\alpha_s}} \bl( \bl( \frac{v'}{v} \br)^3 \ \frac{y_c' y_b' y_t'}{y_c y_b y_t} \br)^{2/27} \bl( \frac{m_c m_b m_t}{\Lambda_\UV^3} \br)^{\frac{2}{27} \delta_{\alpha_s}} 
    \!\!\!\!\!\!\!\!
    \prod\limits_{m_t < m_i < \Lambda_\UV}
    \!\!\!\!
    \bl( \frac{\Lambda_\UV}{m_i} \br)^{\frac{b_i}{9} \delta_{\alpha_s}} \ , \label{eq:lambda_ratio2} \\
    \text{defining} \quad \delta_{\alpha_s} & \equiv & \frac{\Delta \alpha_s}{\alpha_s'} \equiv \frac{\alpha_{s,\UV}' - \alpha_{s,\UV}}{\alpha_{s,\UV}'} \quad \text{for brevity.} \nonumber 
\eea
\Eq{eq:lambda_ratio2} can be obtained from \Eq{eq:lambda_ratio1} by writing $\alpha_{s, \UV}$ in terms of $\Lambda_\UV$ and new physics above the top-quark mass in the form of new colored particles. Of course, since for our scenario to work it is imperative to solve the electroweak hierarchy problem, there may very well be many of these new color states. Throughout the rest of this section we consider the additional color content associated with six-flavored Supersymmetric QCD (SQCD), where the new colored states are squarks and gluinos, and $B_{\rm NP} = -4$. SQCD is of course part of a supersymmetric solution to the electroweak hierarchy problem.

\begin{figure}[h!]
  \centering
  \includegraphics[width=0.7\textwidth]{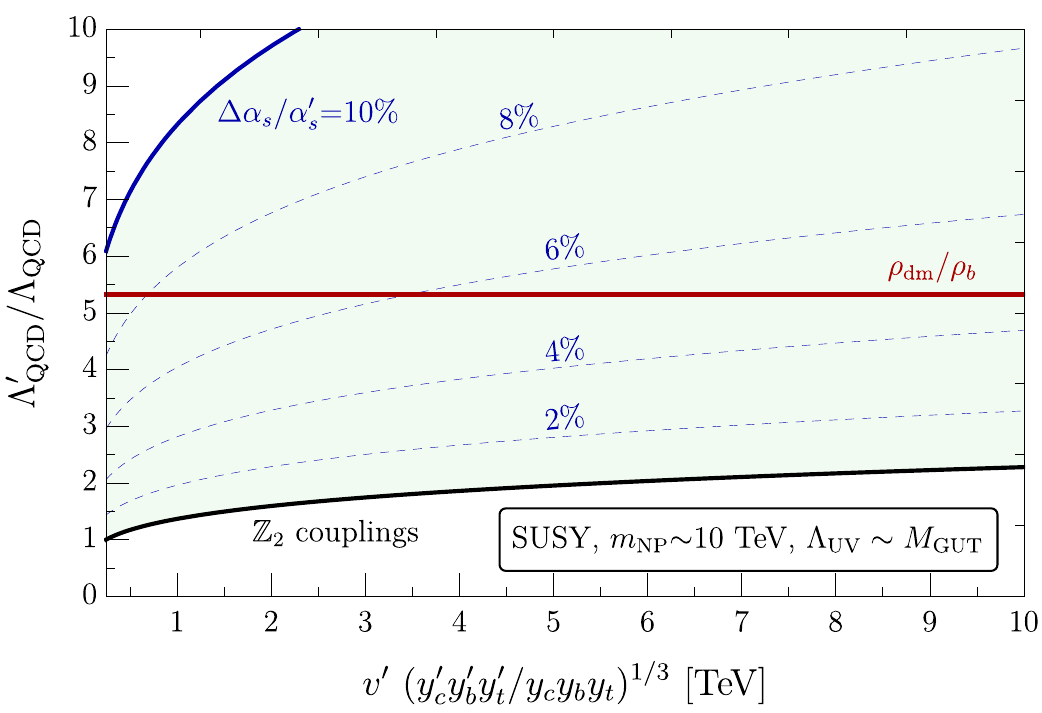}
  \caption{
  $\LQCDp/\LQCD$ ratio for 3 light dark quarks, given by \Eq{eq:lambda_ratio2}, and plotted as a function of $v' \bl( y_c' y_b' y_t' / y_c y_b y_t \br)^{1/3}$. We have chosen SQCD as an illustrative example of the impact that an UV completion has on this ratio. We take $\Lambda_\UV = M_{\rm GUT} \sim 10^{15}~\GeV$, and $m_{\rm NP} \sim 10~\TeV$ as the mass of new colored states, namely the squarks and gluinos. The solid black line corresponds to exact $\Ztwo$-symmetric strong couplings (\ie~$\Delta \alpha_s / \alpha_s' = 0$). The blue lines correspond to various choices of the detuning $\Delta \alpha_s / \alpha_s'$ between the visible and dark strong couplings at the UV scale. The horizontal red line is the target value of $m_{b'} / m_b = \rho_\dm / \rho_b \approx 5.3$.}
  \label{fig:confine}
\end{figure}

In \Fig{fig:confine} we plot the ratio $\LQCDp/\LQCD$ for our model, as a function of $v' \bl( y_c' y_b' y_t' / y_c y_b y_t \br)^{1/3}$, for GUT-scale $\Lambda_\UV$ and new SQCD physics, the squarks and gluinos, with masses $m_i = m_{\rm NP} = 10~\TeV$. We show lines corresponding to $\Ztwo$-symmetric strong couplings ($\Delta \alpha_s / \alpha_s' = 0$), as well as various percentage-level detunings $\Delta \alpha_s / \alpha_s'$ of the visible and dark strong coupling. This figure illustrates that only modest $\Ztwo$-breakings in the dark VEV $v'$, the dark heavy quark Yukawas $y_i'$, and in the strong coupling $\alpha_s'$, are required for $m_{b'} / m_b \simeq \LQCDp / \LQCD$ to reach the target value given in \Eq{eq:rho_ratio_bench}.

As it turns out, $\Delta \alpha_s / \alpha_s' \lesssim 10\%$ detunings are readily obtained in our extra-dimensional scenario. Indeed, the 4D gauge couplings $\altp{\alpha_4}$ are related to their (dimensionful) 5D versions $\altp{\alpha_5}$ by the size $L$ of the extra dimension:
\beq
    \frac{1}{\alpha_4^{(\prime)}} = \tau_{\rm bdy}^{(\prime)} + \frac{L}{\alpha_5} \ ,
\eeq
where $\alpha_5' = \alpha_5$ in the $\Ztwo$-symmetric bulk, and $\altp{\tau_{\rm bdy}}$ represents boundary contributions, which due to the left-hand boundary of \Fig{fig:orbifold} are $\Ztwo$-breaking. If these boundary contributions break $\Ztwo$ at $\mO(1)$ level, a small detuning 
naturally emerges the larger the extra dimension is, relative to the scale characterizing the non-renormalizable $\alpha_5$. 
The requisite
$\Delta \alpha_s / \alpha_s' \lesssim 10\%$ can be achieved if $\alpha_5 / L \lesssim 1/10$.\footnote{For comparison, in orbifold SUSY GUTs the analogous extra-dimensional diluted GUT-breaking among gauge couplings must a few percent in order to fit data and SM two-loop running (see the GUT review in Ref.~\cite{Workman:2022ynf}).}

As explained in the introduction, it is critical in our framework that the electroweak hierarchy problem (in each sector) is broadly solved by new physics, in order for the $\Ztwo$ symmetry to be sufficiently well respected at observable energies. This still allows for a {\it little} hierarchy problem to remain, and the absence of new physics in current data suggests there is indeed such a little hierarchy in the VS between the weak scale and the scale of new physics $\gtrsim {\cal O}(10)$ TeV, at the cost of modest fine-tuning. The DS will have nearly the same scale of new physics by the approximate $\Ztwo$, but by  being ``de-tuned'' with respect to the finely-tuned VS it will have a larger electroweak scale, $v' > v$. Within our extra-dimensional geography, the de-tuning is only (fractionally) significant in the smaller Yukawa couplings. \Fig{fig:confine} shows  $\Delta \alpha_s / \alpha_s' \lesssim 10\%$ detuning of the QCD$^{(\prime)}$ couplings needed in order to achieve the requisite $\sim 5$ GeV dark baryon mass, so we can expect all gauge-couplings detuned at approximately this level. With this level of electroweak gauge coupling detuning, the $W'$-bosons contribution to the dark Higgs self-energy means that we can roughly expect $v' \sim \mO(10) \times v$. Later, in \Sec{subsec:nucl} we will see that this expectation that $v' \sim $ TeV fits nicely with confidently avoiding the formation of complex dark nuclei (and hence dark atoms) in the early Universe, which would be at odds with DM observations.

\subsection{Symmetric Dark Relics}
\label{subsec:symm}

Having discussed the origin of the tiny particle-antiparticle asymmetry in the abundances of the visible and dark baryons, we now need to make sure that this is the {\it only} abundance left. In other words, we need to ensure that the particle-antiparticle {\it symmetric} components annihilate into the light species of their respective sectors sufficiently early so as not to overclose the Universe.

In our scenario, these symmetric components originate from both the primordial bath (the dominant contribution), and from the $(1 - \altp{\eCP}) \altpp{Y}{\chi}$ $CP$-symmetric leftover from the $\altp{\chi}$ decays (the subdominant one, given the smallness of $\altpp{Y}{\chi}$). The most dangerous states are those charged under a conserved quantum number (such as baryon number or electric charge) or light massive particles that may survive into late times. These are the baryons, the pions, and the electrons, as well as their dark counterparts. We remind the readers that the dark neutrinos and dark photons form part of the light relics that constitute the $\DNeff$ we derived in \Eq{eq:DNeff}.

In the VS, of course, the symmetric abundances of all these particles annihilate or decay away in a very efficient manner. Indeed, baryon-antibaryon annihilations, mediated by the strong interactions, produce tiny amounts of relic abundances after they freeze out at temperatures below $\mO(1~\GeV)$ (hence the need for a baryogenesis mechanism!), while $e^+ e^- \to \gamma \gamma$ do likewise below $\mO(1~\MeV)$. The $\pi^+ \pi^- \to \pi^0 \pi^0$ annihilations (also with strong cross sections), together with the extremely prompt (compared to the Hubble time) $\pi^\pm \to \mu^\pm \nu$ and $\pi^0 \to \gamma \gamma$ decays, take care of the symmetric pion abundance, which is then gone shortly after they become non-relativistic.

Things are very much the same for the DS. The fact that the $\Ztwo$ symmetry relating both sectors is very good, and that the dimensionful scales $v'$ and $\LQCDp$ are only somewhat larger than their SM equivalents, guarantees that the same story is repeated play-by-play with the dark particles. This is true despite the fact that they are part of a colder bath, that they interact with typically smaller dark cross sections (due to larger mass scales), and that the dark hadrons and charged leptons are heavier, all of which translate into earlier freezeouts and larger contributions to the total energy density compared to their visible equivalents. Nevertheless, the efficiency with which the visible sector gets rid of its symmetric abundances is so large that our scenario has plenty of breathing room to accommodate modest changes in mass scales without making a problem of the dark symmetric relic abundances. For example, despite their smaller cross section (due to the heavier dark pions mediating the dark strong nuclear force), dark baryon-antibaryon annihilation is so efficient that a symmetric relic abundance of only $\Omega_{b'\anti{b'}} \sim 4 \times 10^{-10}$ (for our benchmark parameters) remains after freezeout.

\subsection{(Absence of) Dark Nucleosynthesis}
\label{subsec:nucl}

As noted before, the dark neutron ($n'$) makes an ideal dark matter candidate as it can readily satisfy the self-interaction constraint on DM from the Bullet Cluster \cite{Markevitch:2003at,Randall:2008ppe,Robertson:2016xjh} (namely, that their self-interacting cross section and mass obey $\sigma_{n'n'} / m_{n'} \approx \pi m_{\pi'}^{-2} / m_{n'} \approx 3 \times 10^{-4}~\cm^2/\gr < 1~\cm^2 / \gr$),\footnote{The magnetic dipole moment cross section $\sigma_{\rm magnetic} \sim \alpha_{\rm EM'}^2 m_{n'}^{-2}$ is much smaller than $\sigma_{n'n'} \sim \sigma_{\rm strong} \sim \pi m_{\pi'}^{-2}$ from strong interactions.} while simultaneously relating the abundances of dark matter and baryons. Dark protons ($p'$), on the other hand, would lead to the formation of atomic dark matter, which has a much larger scattering cross section.\footnote{The relevant constraints are different if these dark atoms are ionized. Indeed, the long-range dark Coulomb interactions between dark nuclei and dark electrons would run afoul a plethora of observations; cooling of the ionized dark gas could take place and dark disks form, which are severely constrained. However, a quick estimate shows that if 10\% of the total DM mass is in ionized dark hydrogen (with a nucleus mass of $m_{H'} = m_{p'} \approx 5~\GeV$ and $m_{e'} \approx (y_e' v'/ y_e v) m_e \approx 4 m_e (v' / 1~\TeV) (y_e' / y_e)$), the {\it Bremsstrahlung} cooling timescale is $t_{\rm cool} \sim 3 \times 10^{10}~\yr$, larger than the age of the Universe \cite{Fan:2013yva}. If the dark atoms have instead an atomic mass $A'$ larger than that of dark hydrogen (\eg, dark tritium or dark helium), the timescale is increases by a factor of $A'$. Therefore, satisfying the Bullet Cluster constraints automatically prevents the formation of dangerous dark disks for most of our model's parameter space.} Such dark atoms can only constitute at most $\sim \mathcal{O}(10\%)$ of DM \cite{Robertson:2016xjh}; out of an abundance of caution we would like to avoid them entirely. In the dark sector, the abundance of $p'$ can be suppressed by making them heavier than $n'$, allowing $p' \to n' e'^{+} \nu_e'$ decays. This is easily accommodated in our model, as the Yukawas of the light up- and down-type quarks are not $\Ztwo$-symmetric. A heavier $p'$, however, is not sufficient to avoid dark atoms, as there could be nucleosynthesis in the DS, analogous to big bang nucleosynthesis (BBN) in the VS. During dark nucleosynthesis (dark BBN) \cite{Berezhiani:1995am,Berezhiani:2005ek,Krnjaic:2014xza,Detmold:2014qqa,Hardy:2014mqa,Hardy:2015boa,Mitridate:2017izz,Gresham:2017cvl,Redi:2018muu,Mahbubani:2020knq}, the otherwise unstable dark protons could be stabilized inside dark nuclei, just as SM neutrons can be stabilized inside SM nuclei. We must therefore carefully consider the conditions necessary for the absence of dark BBN, to avoid the formation of undesirable atomic dark matter.


As in the SM, the dark weak interactions $ p' + e'^{-} \rightleftharpoons n' +  \nu'_{e} $, $p' + \bar{\nu}'_{e} \rightleftharpoons n' +  e'^{+}$, and   $p' \rightarrow n' + e'^{+} + \nu'_{e}$ maintain thermal equilibrium between $n'$ and $p'$. As the temperature drops below the mass difference between the dark nucleons, $\Delta m' = m_{p'}-m_{n'}$, the abundance of $p'$ decreases compared to $n'$, until the relevant dark weak interactions freeze out at temperature $T'_{w{\rm fo}}$. The ratio of number densities of $p'$ and $n'$ at freeze-out is then given by
\beq
    \left(\frac{n_{p'}}{n_{n'}}\right)_{w{\rm fo}} = \exp\left(-\frac{\Delta m'}{T'_{w{\rm fo}}}\right) \ .
\eeq
The fraction of dark nucleons that end up in dark nuclei depends on the amount of $p'$ that remains until the onset of dark BBN and the heaviest dark nuclei that can be successfully assembled from these leftover $p'$. This can be analyzed in close analogy with the standard BBN.

The first nucleus to form is the dark deuterium ($D'$). In the SM, efficient deuterium formation begins at a much lower temperature compared to the deuterium binding energy, $T_{D} \approx E_{D}/20 \approx 0.1~\MeV$, also known as the deuterium bottleneck. A similar bottleneck must also exist for $D'$ formation, $T'_{D'} \approx E_{D'}/20$, which would delay dark BBN. Once $D'$ is formed, it can either capture a $n'$ to make dark tritium ($\mathrm{Tr}'$) or fuse with another $D'$ to make dark helium (${\rm He}'$).\footnote{While in the SM free neutrons are unstable, they are stable inside deuterium because the nuclear binding energy is larger than the neutron-proton mass difference. This, however, is not necessarily true in the DS. Indeed, $\Delta m' > E_{D'}$ for part of our parameter space, rendering $D'$ unstable. This means that the dark deuterium bottleneck is further narrowed by the finite $D'$ lifetime, and in the end smaller amounts of stable dark nuclei will be formed. We will henceforth ignore this subtlety, and conservatively assume that all the originally synthesized $D'$ survive long enough to form heavier dark nuclei.} While ${\rm He}'$ is typically more stable than $\mathrm{Tr}'$ (in the SM, for example, most of the synthesized deuterium converts into helium), the cross section for the fusion of the charged nuclei required to make ${\rm He}'$ and heavier elements is exponentially suppressed due to the Coulomb barrier \cite{Kolb:1990vq}. This is what prevents the creation of a significant abundance of nuclei heavier than Helium in the SM. Due to the larger mass scales and colder temperature in the DS, its nucleosynthesis is less efficient, thereby also preventing the formation of dark nuclei with atomic masses above ${\rm He}'$. In those parts of our parameter space where the Coulomb barrier suppression is not strong enough at the time of $D'$ formation, we expect all of $D'$ (and therefore, all of leftover $p'$) to end up in ${\rm He}'$. On the other hand, in those parts where $D'$ formation is sufficiently delayed due to the low $D'$ binding energy, most of $p'$ will instead end up in $\mathrm{Tr}'$. Out of these two scenarios, the one with mostly $\mathrm{Tr}'$ has the largest mass fraction in dark atoms, since $\mathrm{Tr}'$ assimilates more $n'$ per $p'$ into dark nuclei. For our constraint, therefore, we focus on the more conservative case where all the leftover $p'$  ends up in $\mathrm{Tr}'$ in the entire parameter space. The mass fraction of $\mathrm{Tr}'$ is then
\beq\label{eq:XTr}
    X_{\mathrm{Tr}'} = \left(\frac{3 n_{p'}}{n_{p'}+n_{n'}}\right)_{D'} = \frac{3(n_{p'}/n_{n'})_{D'}}{(n_{p'}/n_{n'})_{D'}+1} \ .
\eeq
Taking into account the decay of $p'$ between freeze-out and dark BBN, the dark proton-to-neutron ratio at the onset of dark BBN can be written as
\beq\label{eq:npnd_Dp}
     \left(\frac{n_{p'}}{n_{n'}}\right)_{D'} = \left(\frac{n_{p'}}{n_{n'}}\right)_{w{\rm fo}} \, \exp\left(-\frac{(t_{D'}-t_{w{\rm fo}})}{\tau_{p'}}\right) \approx \left(\frac{n_{p'}}{n_{n'}}\right)_{w{\rm fo}} \, \exp\left(-\frac{1}{H_{D'} \,\tau_{p'}}\right) \ ,
\eeq
where $t_{D'}-t_{w{\rm fo}}$ is the time difference between the freeze-out and $D'$ formation. We can simplify $(t_{D'}-t_{w{\rm fo}}) \approx t_{D'} $, which can be written in terms of the Hubble constant at $D'$ formation, $ t_{D'}  \sim H^{-1}_{D'}$. Bullet Cluster observations require a the dark atom mass fraction to be less than 10\%; in our conservative estimate this means $X_{\mathrm{Tr}'} \lesssim 0.1$. Plugging \Eq{eq:npnd_Dp} into \Eq{eq:XTr} this bound can be translated into a constraint on the allowed amount of $p'$ remaining by the time of dark BBN, 
\begin{align}
     \left(\frac{n_{p'}}{n_{n'}}\right)_{D'} \approx \left(\frac{n_{p'}}{n_{n'}}\right)_{w{\rm fo}} \, \exp\left(-\frac{1}{H_{D'} \,\tau_{p'}}\right) &\lesssim 0.03 \nonumber \\
    \rightarrow \quad \exp\left(-\frac{\Delta m'}{T'_{w{\rm fo}}}\right)  \exp\left(-\frac{1}{H_{D'} \,\tau_{p'}}\right) &\lesssim 0.03 \ . \label{eq:allowed_p'_abundance}
\end{align}

\begin{figure}[bt!]
  \centering
  \includegraphics[width=0.6\linewidth]{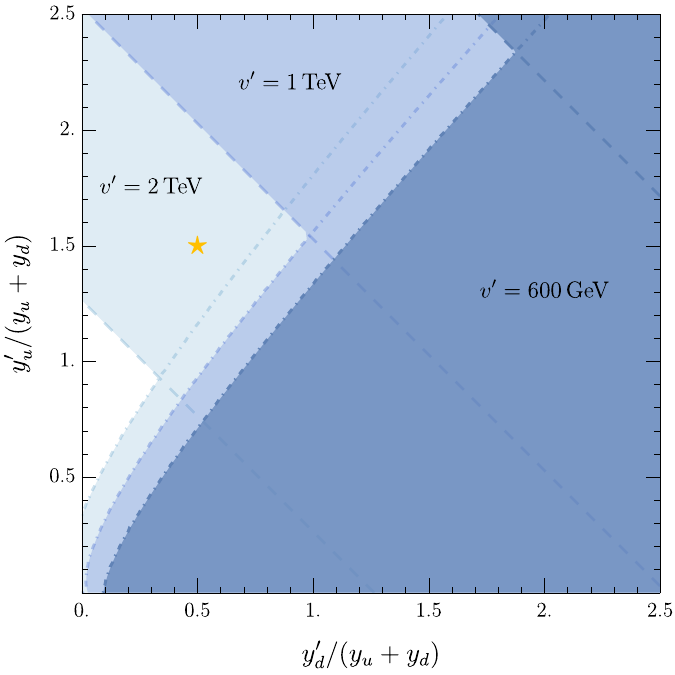}
  \caption{
  Constraints on the parameter space of $y_u'$ and $y_d'$, where the blue shaded regions are excluded by the consideration of forming atomic dark matter. The shades, from darker to lighter, correspond to three choices of increasing dark weak scale, $v'=600$ GeV, $v'=1$ TeV, and $v'=2$ TeV, respectively. The regions bounded by dot-dashed lines are excluded due to high abundance of leftover dark protons and correspond to the constraint in \Eq{eq:allowed_p'_abundance}, while the dashed lines enclose excluded regions that correspond to the constraint on dark dineutron stability given in \Eq{eq:pion_baryon_ratio}. We have fixed $\LQCDp/\LQCD =5.3$ and $\xi=0.44$. The benchmark considered in the text, $y'_{u} = 1.5 \, (y_u +y_d) \approx 6 \times 10^{-5}$ and $y'_{d} = 0.5 \, (y_u +y_d) \approx 2 \times 10^{-5}$ for $v' = 1$ TeV, is represented by a yellow star.}
  \label{fig:nodBBN}
\end{figure}

It will be useful to evaluate various parameters in \Eq{eq:allowed_p'_abundance} in terms of their SM counterparts. The mass difference $\Delta \altp{m}$ contains two contributions: one from the quark masses and the other from the electromagnetic contribution to proton mass, $\Delta \altp{m} = \Delta \altp{m}_{\rm quark} + \Delta \altp{m}_{\rm EM}$. In the SM, $\Delta m_{\rm quark} \approx 2.3~\MeV$ and $ \Delta m_{\rm EM} \approx -1~\MeV$ \cite{BMW:2014pzb,Brantley:2016our,FlavourLatticeAveragingGroupFLAG:2021npn} have opposite signs, and partially cancel to give a smaller $\Delta m \approx 1.3~\MeV$ \cite{Gasser:2020mzy, Workman:2022ynf}. In the DS, however, these two terms will add up, since $p'$ is heavier. The two contributions can then be estimated from those in the SM by simple rescalings of the fundamental parameters. Indeed, $\Delta m'_{\rm quark} \approx \Delta m_{\rm quark} \bl( \frac{y'_{u}-y'_{d}}{y_d - y_u} \br) \bl( \frac{v'}{v} \br)$ and $\Delta m'_{\rm EM} \approx (\Lambda'_{\rm QCD}/\Lambda_{\rm QCD})\, 1~\MeV$. Consider a benchmark of $v' = 1~\TeV$ and $\LQCDp / \LQCD = \rho_\dm / \rho_b \approx 5.3$ with $y_u' = 1.5 (y_u + y_d) \approx 6 \times 10^{-5}$ and $y_d' = 0.5 (y_u + y_d) \approx 2 \times 10^{-5}$ (denoted by the yellow star in \Fig{fig:nodBBN}), where these mass differences amount to $\Delta m'_{\rm quark}  \approx 25~\MeV$ and $\Delta m'_{\rm EM} \approx 5.3~\MeV$, giving a total mass splitting of $\Delta m' \approx 30~\MeV$. 

The freeze-out temperature $T'_{w{\rm fo}}$ can be estimated by comparing the rate of interaction to Hubble, $\Gamma \sim G_{F}^{\prime 2} T^{\prime 5} \sim H \sim T^{2}/M_{\Pl}$, where $G^{(\prime)}_{F}$ is the (dark) Fermi constant. Note that by this time, temperature of the DS is smaller than the SM ($T' < T$) due to asymmetric reheating discussed in \Sec{subsec:asymmReheat}. Let us denote the temperature ratio between the two sectors by $\xi \equiv T'/T$. $\xi \leq 0.44$ is required to satisfy the $\DNeff$ constraints. Using $G^{(\prime)}_{F} \propto 1/v^{(\prime)2}$, the $T'_{w{\rm fo}}$ can be related to the corresponding temperature in the SM, $T_{w{\rm fo}} \approx 1$ MeV, as
\begin{align}\label{eq:temp_p'_freezeout}
    \frac{T'_{w{\rm fo}}}{T_{w{\rm fo}}} \approx \left(\frac{v'}{v}\right)^{4/3} \frac{1}{\xi^{2/3}},
\end{align}
assuming $\alpha_{\rm EM} \approx \alpha'_{\rm EM}$. For the benchmark $v'=1$ TeV and $\xi =0.44$, we get $T'_{w{\rm fo}} \sim 11$ MeV. 

As mentioned before, the time of $D'$ formation is related to the $D'$ binding energy as $T'_{D'} \approx E_{D'}/20$. Therefore,
\beq\label{eq:temp_D'_formation}
    T'_{D'} \sim \frac{E_{D'}}{E_D} T_{D} \,.
\eeq
The ratio of $E_{D}$ to nucleon mass $m_n$ must scale as \cite{NPLQCD:2012mex}
\beq
    \frac{E_{D}}{m_n} \propto \left(\frac{m_{\pi}}{m_{n}}\right)^{2} \,.
\eeq
Therefore, $E_{D'}$ can be determined using the relation
\beq\label{eq:BE_D'}
     E_{D'} \approx 
     \frac{m_{\pi'}^{2} \, m_{n}}{m_{\pi}^{2} \,m_{n'}} \, E_{D} \approx \frac{(y'_{u}+y'_{d}) \, v'}{(y_{u}+y_{d}) \, v} \, E_{D} \,,
\eeq
where $E_D \approx 2.2~\MeV$ in the SM.
For the benchmark from before with $v' \approx 1~\TeV$, $\xi \approx 0.44$, and $(y'_{u}+y'_{d})/(y_{u}+y_{d}) \approx 2$, we get  $E_{D'} \sim 20~\MeV$, and $T'_{D'} \sim 0.9$ MeV with the SM temperature $T_{D'} \approx 2$ MeV. The temperature of $\sim 1$~MeV in the SM corresponds to $H^{-1}|_{1\,{\rm MeV}} \approx 1$~sec. Therefore, for a general $T_{D'}$
\beq\label{eq:hubble_D'}
    H_{D'} \approx \left(\frac{T_{D'}}{1\,{\rm MeV}}\right)^{2} (1~{\rm sec})^{-1} .
\eeq
It is also useful to cast the lifetime of dark protons in terms of the lifetime of SM neutrons,
\bea
    \frac{\tau_{p'}}{\tau_{n}} \!\!\!& \approx &\!\!\! \frac{G_{F}^{2} \, m_{e}^5 \, f_{p}(\frac{\Delta m}{m_{e}})}{G_{F}^{\prime 2} \, m_{e}^{\prime 5} \, f_{p}(\frac{\Delta m'}{m_{e'}})}  \approx  1.8 \times 10^{-5} \ \bl( \frac{1~\TeV}{v'} \br) \bl( \frac{y_e}{y_e'} \br)^5 \bl( \frac{2.3 \times 10^4}{f_p(\frac{\Delta m'}{m_{e'}})} \br) \nonumber \\ 
    && \hspace{5.5em} \xrightarrow{{\tiny \Delta \altp{m} \, \gg \, \altp{m_e}}} \hspace{0.1em} 1.8 \times 10^{-5} \ \bl( \frac{v'}{1~\TeV} \br)^{4} \bl( \frac{30~\MeV}{\Delta m'} \br)^5 \, , \label{eq:lifetime_rescale}
\eea
where $f_{p}(x) = \frac{1}{60}(2x^{4}-9 x^{2} - 8)(x^{2}-1)^{1/2}+\frac{1}{4}x \log(x+(x^2-1)^{1/2})$ is the phase space factor \cite{WILKINSON1982474}\footnote{In the SM, $f_p(\Delta m/m_e) \approx 1.6$. As we demonstrate shortly, the argument $x = \Delta m' / m_e'$ in the DS is large, and thus $f_p(x) \approx x^5/30$. We have used this simplified expression for $f_p(x)$ in the last equality of \Eq{eq:lifetime_rescale}.} and $\altp{m_e} = \altp{y_e} \altp{v}/\sqrt{2}$ is the (dark) electron mass in terms of its Yukawa coupling.\footnote{In principle, the small electron-Yukawa coupling may not be $\Ztwo$-symmetric in our setup. However, for the sake of simplicity, we will take $y_{e} \approx y'_{e}$.}

\Eq{eq:allowed_p'_abundance} together with Eqs.~(\ref{eq:temp_p'_freezeout} - \ref{eq:lifetime_rescale}) gives us a constraint on $p'$ abundance that satisfies Bullet Cluster observations. For the benchmark we have been considering above (shown as the yellow star in Fig.~(\ref{fig:nodBBN})), the fraction of leftover dark protons $(n_{p'}/n_{n'})_{D'} \approx 10^{-14}$ easily satisfies the Bullet Cluster bound.

The absence of dark protons is not sufficient to prevent the formation of atomic dark matter. If two $n'$s can form a bound state (dark dineutron), bigger nuclei of pure $n'$ can form during dark BBN, and can grow to be very large due to the lack of electromagnetic repulsion. Such nuclei of pure $n'$ are perfectly valid dark matter candidates. However, as the nucleus grows, the Fermi energy from Pauli blocking increases \cite{Bertulani:2007bfy}, contributing negatively to the binding energy.\footnote{This is the reason behind the sign of the Fermi-energy term of the Bethe–Weizs\"{a}cker semi-empirical formula for the masses of atomic nuclei.} At some point, it will be energetically favorable for the $n'$ inside the nucleus to decay to heavier $p'$ to reduce Fermi energy, and transition to a mixed nucleus of $n'$ and $p'$. This $\beta'$ decay is analogous to $\beta^+$ decay in the SM, in which a proton inside a nucleus decays into a neutron emitting a positron and a neutrino. The stability of dark dineutron can therefore lead to atomic dark matter {\it even if we start with a purely $n'$ plasma prior to dark BBN}. Fortunately, we know that in the SM, the dineutron state is not bound \cite{Spyrou:2012zz}. Lattice simulations \cite{Yamazaki:2015asa, Orginos:2015aya, Yamazaki:2012fn} show that the dineutron state is stable for pion masses $\geq 300$ MeV.\footnote{The lattice simulations of dibaryon bound states involve potentially large systematic uncertainties. For example, different approaches employed in Refs.~\cite{Berkowitz:2015eaa,Wagman:2017tmp} and Ref.~\cite{Amarasinghe:2021lqa} for $m_{\pi}\sim 800$ MeV seem to produce different conclusions about the stability of dineutron bound state. While the results for $m_{\pi} \sim 300$ MeV that we have used in this paper are currently undisputed, our conclusions are subject to further verification of these results with improved lattice simulations.} The stability of dineutron state is unclear in the intermediate mass range $ 135 \, \MeV < m_{\pi} < 300 \, \MeV$ due to the lack of lattice simulations. Since the negative binding energy of the dineutron state in the SM is quite small, we impose a relatively conservative condition on the dark pion mass compared to the lattice results to avoid dark dineutron bound state:
\beq \label{eq:pion_baryon_ratio}
    \frac{m_{\pi'}}{m_{n'}} < 0.2 .
\eeq
Finally, even when dark dineutron is not a bound state, we note that it is crucial that dark BBN stops at $\mathrm{He}'$.  If it were allowed to go to higher atomic number, eventually $\beta'$ decay, combined with dark neutron capture, would form an efficient mechanism of increasing the abundance of dark protons and therefore potentially increasing the abundance of atomic dark matter. But at low atomic numbers, such as Tr$'$ and He$'$, $\beta'$ decay does not occur, analogous to the absence of $\beta^+$ decay in SM Tritium and Helium.

\Fig{fig:nodBBN} shows the constraints  \Eqs{eq:lifetime_rescale}{eq:pion_baryon_ratio} on the parameter space of dark up- and down-type Yukawas. We have fixed $\Lambda'_{\rm QCD}/\Lambda_{\rm QCD} =5.3$  and $\xi=0.44$ in Fig.~(\ref{fig:nodBBN}). The blue shaded regions are excluded due the formation of atomic dark matter. The shades, from darker to lighter, correspond to three choices of increasing dark weak scale, $v' = 600 ~\GeV$, $1~\TeV$, and $2~\TeV$, respectively. For clarity, the regions excluded by  \Eq{eq:allowed_p'_abundance} and \Eq{eq:pion_baryon_ratio} are bounded by dot-dashed and dashed lines respectively. From Fig.~(\ref{fig:nodBBN}), we can see that for $v' \lesssim (\text{few})~\TeV$, dark BBN can be avoided with ${\cal O}(1)$ differences between the VS and the DS values of the Yukawas of the first generation. However, as we take $v'$ to be larger, the viable parameter space is squeezed. Because of this, $v'$ cannot be much larger than $v$, further motivating our benchmark of $v' \sim 1~\TeV$. This scale for $v'$ is in harmony with the discussion of the little hierarchy problem in \Sec{sec:intro}, as well as with the expectation from $10\%$ detunings in the gauge couplings within our extra-dimensional setting, as laid out in \Sec{subsec:confine}.

\section{Phenomenology}
\label{sec:pheno}

Above, we have detailed our dark-sector/visible-sector philosophy, (approximate) symmetry, model realization, and cosmic history. We have seen that the $\DNeff$ bound on new light species strongly constrains this framework, and therefore requires asymmetric reheating, which can readily take place if there is a period of early matter (reheaton) dominance and a moderate hierarchy between the dark and visible electroweak scales due to their high sensitivity to $\Ztwo$-breaking. Of course, it is important to discuss the ways in which we can hope to experimentally detect the dark-sector particle physics beyond just the gravitational collective effects of dark neutrons as CDM. 

Naively, non-gravitational interactions between the visible and dark sectors are dangerous because even relatively weak couplings would have enough time to equilibrate the two sectors to the same temperature, thereby running afoul of the $\DNeff$ constraint. However, this danger is only realized if the inter-sector couplings are large enough at temperatures below that of reheating $T_\rh \approx 70~\MeV$ (see \Eq{eq:Trh_appx}), so low because it reflects the long decay-time needed for the reheaton to dominate the energy density of the Universe. Therefore inter-sector interactions can be relatively strong at high energies if they involve heavy particles, which largely decouple at low energies and temperatures consistently with $\DNeff \ll 1$. This provides two experimental opportunities: {\it (i)} high-energy colliders should be sensitive to the heavy-particle inter-sector physics, and {\it (ii)} dark matter direct detection is also feasible despite the low momentum-exchange with nuclei, and hence very weak interactions with them, because the weak couplings can be offset by the large volume of nuclei, flux of dark matter (dark neutrons), low backgrounds, and long running times. 

To somewhat more quantitatively explore the possibility of inter-sector couplings, consider a generic heavy mediator of mass $M$ bridging the two sectors with a coupling $g$. Scatterings involving relativistic particles from both thermal baths can put them in thermal contact. The thermally-averaged cross section associated with these processes will be $\langle \sigma_{\rm portal} v \rangle \sim g^4 T^2 / (64 \pi M^4 )$. Conservatively demanding that the corresponding scattering rate $\Gamma_{\rm portal} = n \langle \sigma_{\rm portal} v \rangle$ be smaller than the Hubble expansion rate at the beginning of reheaton domination ($T_{R\eq} \sim 1~\GeV$ for $m_R = 700~\GeV$), we arrive at the constraint
\beq
    \frac{M}{g} \gtrsim 5~\TeV \ \bl( \frac{75}{g_{*R\eq}} \br)^{1/8} \bl( \frac{200}{g_{*sR\fo}^\tot} \br)^{3/4} \bl( \frac{m_R}{700~\GeV} \br)^{3/4} \ . \label{eq:portal_decoupling}
\eeq
Thus, we can indeed have roughly (multi-)TeV mediators with $\lesssim {\cal O}(1)$ couplings, consistent with the dark sector remaining much colder than the visible sector after reheating by reheaton decay. Below, we will illustrate this general possibility with two elegant and natural examples.

\subsection{The Higgs Portal}

The first example is the ``Higgs portal'', the possibility of a renormalizable quartic coupling between the visible and dark Higgs fields of the form $\lambda_{HH'} \abs{H}^2 \abs{H'}^2$. Such a $\Ztwo$-symmetric coupling can be thought of as localized on the $\Ztwo$-symmetric boundary of our extra-dimensional realization. At low temperatures/energies neither Higgs particle can be on-shell but they mediate an effective coupling between the fermions in the thermal baths of the visible and dark sector as $g^2 \sim \lambda_{HH'} y_f y_f'$, while the effective ``mediator'' mass is $M^2 \sim (m_h^2 m_{h'}^2) / (v v')$. In this case, the constraint in \Eq{eq:portal_decoupling} states that
\beq
    \lambda_{HH'} < 11 \ \bl( \frac{g_{*R\eq}}{75} \br)^{1/4} \bl( \frac{g_{*sR\fo}^\tot}{200} \br)^{3/2} \bl( \frac{700~\GeV}{m_R} \br)^{3/2} \bl( \frac{v'}{1~\TeV} \br) \bl( \frac{y_c}{y_f'} \br) \qquad (\text{for } c \anti{c} \to f' \anti{f'} \text{ scattering}) \ ,
\eeq
a very loose bound indeed. Here $y_c$ is the SM charm quark Yukawa, which is the fermion involved in these processes at a $T_{R\eq}$ temperature of around $1~\GeV$. 

In our minimal model the reheaton already acts as one such Higgs-dark Higgs portal, integrating it out induces a ridiculously small and uninteresting coupling $\lambda_{HH'} = \beta_R \beta_R' / m_R^2 \approx 2 \times 10^{-22} ( \beta_R / 10~\eV ) ( \beta_R' / 10~\eV ) ( 700~\GeV / m_R )^2$. However, a renormalizable Higgs portal can clearly be much stronger. In this case, the central constraint on $\lambda_{HH'}$ is not cosmological in nature, but rather comes from its impact on precision Higgs physics. After electroweak and dark-electroweak symmetry breaking this coupling induces mixing between the Higgs and dark Higgs bosons so that the experimental $125~\GeV$ mass eigenstate has a small admixture of non-SM $H'$. This leads to an overall reduction in its couplings to other SM particles, relative to SM predictions. In our $v' = 1~\TeV$ benchmark, current Higgs measurements imply~\cite{Ciuchini:2013pca,deBlas:2016nqo,Contino:2017moj}
\beq
    \lambda_{H H'} \lesssim 0.23 \bl( \frac{v'}{1~\TeV} \br).
\eeq
Yet, the related deviations in Higgs couplings could still be discovered at the high-luminosity LHC or at future colliders~\cite{deBlas:2019rxi}. With sufficient energy one could even create dark sector states via this portal coupling, escaping as missing energy. 

Famously, there is a special value of the portal coupling, $\lambda_{H H'} \approx 2 \lambda_{\rm SM}$ (with $\lambda_{\rm SM} \approx 0.13$ the SM Higgs quartic coupling), where the Higgs potential of both sectors $V(H, H')$ has an enhanced approximate $SU(4) \supset SU(2)_{\rm EW} \times SU(2)_{\rm EW'} $ symmetry. This enhanced symmetry is the core of the Twin Higgs mechanism for ameliorating the little hierarchy problem \cite{Chacko:2005pe,Chacko:2005un,Chacko:2005vw,Burdman:2014zta,Craig:2015pha}. For this mechanism to operate in the $\sim \TeV$ regime, the enhanced approximate symmetry has to be enforced by the UV completion of Twin Higgs that solves the larger hierarchy problem. Examples of such UV completions are given in Ref.~\cite{Craig:2013fga} in a SUSY framework and in Ref.~\cite{Geller:2014kta} in a warped extra-dimensional framework.

Here, we do not commit to the Twin Higgs paradigm, pursuing instead our single goal of a natural theory of dark matter and matter. But it is indeed possible that the mystery of dark matter is intimately tied to the solution of the little hierarchy problem in which the dark sector plays a central role, as has been pursued in a number of papers already; see for example Twin Higgs or ``neutral naturalness'' more generally (\cite{Batell:2022pzc} and references therein). We leave for future work whether this can be done consistently with our own check-list of requirements for dark sector modeling. 

\subsection{Massive $B-L$ Gauge Boson Portal}\label{subsec:bl}

The Higgs portal directly couples the two sectors, but we can also consider the possibility of new particles that mediate interactions between the two sectors with appreciable couplings. These new particles and couplings should be consistent with the host of precision flavor, $CP$ and electroweak measurements we already have. In this regard, new massive abelian gauge bosons are particularly safe. While these are usually referred to generically as $Z'$, here we will refer to them as $\hat{Z}$ since we have associated the prime symbol with the dark sector. 

We will consider gauging the anomaly-free (with the addition of three right-handed neutrinos and taking neutrino masses to be predominantly Dirac in nature \cite{Agashe:2005vg}) $U(1)_{B-L}$ symmetry of the SM and the dark sector. We take its gauge field to be $\Ztwo$-invariant, so that it gauges both the visible and dark $U(1)_{B-L}$ with equal strengths. That is, visible and dark baryons/leptons have gauge charges $\pm 1$ under this $U(1)_{B-L}$. Since we want the gauge field $\hat{Z}$ to be massive, we take $U(1)_{B-L}$ to be Higgsed by a $\Ztwo$-invariant scalar field. Given that there is a major piece of new physics we expect at the $\sim 10$ TeV scale responsible for solving the electroweak hierarchy problem(s), such as SUSY, it is natural that this scalar acquires a multi-TeV VEV. Indeed, taking $g_{B-L} \lesssim 1$, implies $M_{\hat{Z}} \lesssim \text{multi-}\TeV$. Once again, the general estimate of \Eq{eq:portal_decoupling} says that such a mediator is cosmologically safe in terms of not re-equilibrating the two sectors after reheating by the reheaton. 

One important consideration is that the WIMP baryogenesis mechanism we employ
relies on breaking baryon-number ($B$), and this must be made consistent with the $U(1)_{B-L}$ gauge dynamics. We assign the baryon parents $\altp{\chi}$ a $B-L$ charge $+1$, and cancel the $B-L$  anomaly introduced by these Weyl fields by assigning charge $-1$ to the accompanying $\altp{\psi}$ fields. Such an assignment makes the decay couplings of the $\altp{\psi}$ to  diquark scalars $\altp{\phi}$ and (dark) quarks $(B-L)$-invariant, consistent with them being ${\cal O}(1)$ in strength. However, the analogous decay couplings of the  baryon parents $\altp{\chi}$ actually violates $B-L$ by charge $2$. This is easily resolved if there is a charge $-2$ ($\Ztwo$-symmetric) Higgs field ${\cal H}$ for $U(1)_{B-L}$, so that we can write the dimension-$5$ operator ${\cal H} \altp{\chi} \altp{\phi} \altp{q}$, which will then lead to a small decay coupling for $\altp{\chi}$ upon ${\cal H}$ acquiring a VEV. For example, if this operator is suppressed by a far-UV scale $\sim 10^{15} $ GeV, we get an effective decay coupling $\epsilon \sim 10^{-11}$, naturally explaining the long lifetime of the baryon parents (see \Eq{eq:epsilon_bounds}). The Majorana mass terms $\altpp{m}{\chi} \chi^{(\prime)2}$ and $\altpp{m}{\psi} \psi^{(\prime)2}$ also break $B-L$, but naturally arise from invariant Yukawa couplings, $ {\cal H} \chi^{(\prime)2}$, and ${\cal H}^* \psi^{(\prime)2}$, upon ${\cal H}$ condensation.

We illustrate how  the phenomenology can be affected with two benchmarks, one with a somewhat lighter and more weakly coupled $\hat{Z}$, and one with a heavier, more strongly coupled $\hat{Z}$. For example, if $M_{\hat{Z}} \sim 300~\GeV$ and $g_{B-L} \sim 0.03$, $\hat{Z}$-exchange between dark matter and nuclei mediates a classic WIMP direct detection signal, the spin-independent scattering cross-section~\cite{Alves:2015pea,Mohapatra:2019ysk}, 
\beq
    \sigma_{\rm SI} = \frac{g_{B-L}^4 m_{\rm nucleon}^2}{\pi M_{\hat{Z}}^4} \approx 10^{-44}~\cm^2 \ \bl( \frac{g_{B-L}}{0.03} \br)^4 \bl( \frac{300~\GeV}{M_{\hat{Z}}} \br)^4 \ ,
\eeq
being at the current exclusion bound~\cite{XENON:2020gfr}. Because the neutrino floor ~\cite{Billard:2013qya} is very close to the exclusion bound for $5$ GeV DM, smaller $g_{B-L}/M_{\hat{Z}} < 1/(10~\TeV)$ will almost always give signals below it. However, there are ongoing efforts to devise experimental strategies to penetrate this neutrino ``fog'' and thereby  access  such small signal cross-sections \cite{OHare:2021utq,Akerib:2022ort}. While such light $\hat{Z}$ are kinematically within LHC reach,  couplings $g_{B-L} < 0.03$ are too small to be currently excluded $\hat{Z}$ \cite{ATLAS:2019erb,EXOCMS,CMS:2021ctt,Workman:2022ynf}. Alternatively, if $M_{\hat{Z}} \sim 30~\TeV$ and $g_{B-L} \sim 0.2$, the on-shell $\hat{Z}$ may be produced at future high-energy colliders. Furthermore, its exchange can mediate production of dark sector states and baryon-parent WIMPs. But while promising for future colliders this benchmark gives a direct detection cross-section orders of magnitude below the neutrino floor, and therefore unlikely to be observable. 
 
The $B-L$ coupling of the baryon-parent WIMP $\altp{\chi}$  provides a new annihilation channel via $\hat{Z}$-exchange in the early Universe. But for the above benchmarks, the associated annihilation cross-sections are subdominant to those arising from $\altp{S}$-exchange, so our $\altp{\chi}$ freezeout and baryogenesis estimates are unaffected. It is an interesting question whether one can simultaneously have $\hat{Z}$ mediate dark-matter direct detection at or not too far below the neutrino floor and provide an observable high energy collider portal to the dark sector. In our current model, we are unable to find such a point of parameter space because of the constraint of not spoiling $\altp{\chi}$ freezeout, but we do see alternative regimes of WIMP baryogenesis in which this should be possible. We leave further variant and/or non-minimal model-building and associated phenomenological implications for future work.
 
%
%


\subsection{Photon-Dark-Photon Mixing Portal}

The massive Higgs and $B-L$ portals we discussed are attractive because they mediate observable interactions while not contributing to low-temperature/energy equilibration of the dark and visible sectors after reheaton-decay,   which would conflict with the observed $\DNeff \ll 1$. By contrast there is a possible photon-dark-photon portal that can mediate dangerous equilibration of just this type, since the photons in both sectors are massless, allowing visible electrons to annihilate into dark electrons which in turn can create other dark species with low mass. Such a portal can arise from hypercharge-dark hypercharge mixing in the far UV, $F_{Y, \mu \nu} F_{Y'}^{\mu \nu}$. Therefore we will take this coupling to vanish in the far UV, which is quite plausible and technically natural. However, it is still important that this coupling is not induced by running down to observable energies, or if it is, that it is naturally very highly suppressed. 

Of course, such an IR-relevant portal between the sectors can only be induced by UV portals. The minimal such connection between the two sectors is the reheaton itself, but it has such a tiny dimensionless coupling, $\beta_R/m_R \sim 10^{-11}$, that it poses no threat. However, in our non-minimal but phenomenologically exciting Higgs and $B-L$ portal scenarios, we must check that they do not induce $F_{Y, \mu \nu} F_{Y'}^{\mu \nu}$ at loop level. For dark charges represented by dark electrons with masses $\sim {\cal O}(1-10)$ MeV, the induced coupling must be $\lesssim 10^{-9}$~\cite{Vogel:2013raa,Chang:2016ntp}. Fortunately, it is straightforward to check that $F_{Y, \mu \nu} F_{Y'}^{\mu \nu}$ is not generated by either portal up to $3$-loop order, which should be adequate suppression. Thus the heavy portals we have used to illustrate the possibility of experimentally probing the dark sector non-gravitationally are safe from the dangerous photon mixing portal.

\subsection{Phenomenology of the Visible Sector}\label{subsec:vs_pheno}

Remaining just in the visible sector, we retain the physics and signals associated with the baryon-parent WIMP, $\chi$ \cite{Cui:2012jh}. If produced at colliders, in particular via $\hat{Z}$-mediated production, it will have potentially long-lived (\ie, displaced-vertex) decays into SM states with net baryon number $\pm 1$ \cite{Cui:2014twa}, symptomatic of it needing to live long enough in the early Universe for it to freeze out before decaying. In addition, since we need to solve the hierarchy problem in each sector, we will be able to discover the associated new physics at multi-TeV high energy colliders. 

Low-energy $\Delta B = 2$ processes, such as baryon-antibaryon oscillations and dinucleon decays, also present potential discovery channels (with $\Delta B = 1$ processes such as proton decay being exponentially suppressed in our model because lepton number remains conserved perturbatively). These are controlled by the $\kappa_i$ and $\lambda_{ij}$ couplings in \Eq{eq:Veach}, which can be very small for the first two generations \cite{Cui:2012jh}. Furthermore, the necessity of flavor antisymmetry in the $\lambda_{ij} \phi d_i^{c \dagger} d_i^{c \dagger}$ term due to color antisymmetry and the anticommutation of the quark fields ensures $\lambda_{dd} = 0$, which in turn means that neutron-antineutron oscillations are highly suppressed. Nevertheless, with only the suppressions associated with Partial Compositeness (namely, extra-dimensional wavefunctions of the light quarks on the boundary supporting $\psi$ and $\phi$) other processes, such as $pp \to K^+ K^+$ dinucleon decays, may lie just beyond current bounds \cite{Workman:2022ynf,Super-Kamiokande:2014hie,Keren-Zur:2012buf,Goity:1994dq,Csaki:2013jza,Super-Kamiokande:2015jbb}. However, such observable signals are not a firm prediction because the $\lambda_{ij}$ may naturally have a substantial overall suppression beyond just the dictates of Partial Compositeness, the only (extremely mild) requirement being that the largest $\lambda_{ij}$ be sufficiently large for the $\phi$ diquarks to decay before BBN.\footnote{By contrast, we do not entertain additional suppressions of the $\kappa_i$'s, since they would in turn further reduce the $\eCP$ asymmetry which is itself a prerequisite of WIMP baryogenesis (see \Eq{eq:eCP}).}

\subsection{$\DNeff$}

The various constraints on the parameter space of our model also correspond to potential discovery channels if experiments improve in these directions. The central such constraint is on the number of new relativistic dark particles, $\DNeff$, in our case represented by the densities of dark photons and dark neutrinos. As shown in \Fig{fig:arh}, our model satisfies the existing Planck satellite 95\% C. R. constraint of $\DNeff < 0.284$ \cite{Planck:2018vyg} by asymmetrically reheating the visible sector via the reheaton decay. But $\DNeff$ cannot be arbitrarily suppressed because the suppression relies on the early matter domination of the reheaton prior to its asymmetric decay (kinematically due to $v' > v$). This early matter dominance is cut off by the reheaton decay, which must occur early enough not to interfere with the physics of BBN and neutrino decoupling. For our benchmark $v' = 1~\TeV$, this gives a minimal $\DNeff > 0.02$. This presents a plausible challenge for upcoming precision cosmological measurements \cite{Dvorkin:2022jyg}, such as the Simons Observatory \cite{SimonsObservatory:2018koc}, CMB-S4 \cite{Abazajian:2019eic}, CMB-HD \cite{Nguyen:2017zqu,Sehgal:2019nmk,Sehgal:2019ewc,Sehgal:2020yja}, MegaMapper \cite{Schlegel:2019eqc}, and PUMA \cite{PUMA:2019jwd}.

\subsection{Other relics}

As we have seen, it is crucial that the $\Ztwo$ symmetry between the visible and dark sectors be broken in the light quark Yukawa couplings, in order for the dark neutron to be the lightest dark baryon and thus the dark matter. Indeed, in our extra-dimensional scenario, inspired by Partial Compositeness, we expect $y_i \neq y_i'$. Beyond the requirements discussed in the introduction (namely, that $y_u' > y_d'$, and that there be three light dark quark flavors), the exact values of the Yukawa couplings are not terribly important. However, if the dark quark strange were to be {\it significantly} lighter than $\LQCDp$, the dark QCD phase transition could be first-order \cite{Fodor:2001pe}. As a result, additional relics such as gravitational waves (generated as the bubbles of the new phase expand and collide) and exotic dark quark matter \cite{Witten:1984rs} could be produced as dark hadrons are formed (\eg, Ref. \cite{Rosenlyst:2023tyj}). We cannot resist pointing out that $\LQCDp \approx 5 \LQCD \approx 1.5~\GeV$ is an energy scale tantalizingly close to the percolation temperature that a first-order phase transition would need to have in order to explain the low-frequency gravitational wave background recently observed by the NANOGrav collaboration \cite{NANOGrav:2023gor,NANOGrav:2023hvm} (for work along these lines see, for example, Refs.\cite{Addazi:2023jvg,Bringmann:2023opz,Fujikura:2023lkn,Jiang:2023qbm}). We leave the study of the phenomenology of such a scenario to future work, and limit ourselves to a SM-like dark sector, where the dark strange quark mass is not too far below the $\LQCDp$, thereby preventing the dark QCD phase transition from being first-order.

\section{Conclusions}
\label{sec:concl}

We have combined bottom-up and top-down considerations to argue that the scenario of a dark sector approximately $\Ztwo$-symmetric with the standard model offers an attractive explanation for the near cosmic coincidence of dark matter and matter. Noting the challenges in the far UV to realizing such a scenario, we presented a simple extra-dimensional framework to produce the requisite hierarchical structure of couplings and $\Ztwo$-breaking, incorporating the mechanism of WIMP (dark) baryogenesis. We also pointed out that the electroweak hierarchy problem must be solved by new physics in order to not 
excessively break the $\Ztwo$ symmetry. 

The strong $\DNeff$ constraint on the abundance of dark relativistic particles represents a central phenomenological challenge. We studied  the case of a long-lived scalar ``reheaton'' field (remnant of the physics solving the hierarchy problem) which can naturally lead to a stage of early matter dominance ending in asymmetric reheating of the two sectors, so as to adequately suppress $\DNeff$. We showed that, to avoid interfering with neutrino decoupling and BBN, $\DNeff \gtrsim 10^{-2}$, which remains within reach of upcoming precision cosmological probes. We also illustrated how heavy ${\cal O}({\rm TeV})$  particles could mediate interactions between the two sectors consistent with suppressed $\DNeff$, giving rise to signatures at high energy colliders as well as dark matter direct detection experiments (albeit most likely lying below the neutrino floor and therefore requiring new experimental methods to uncover). Such ``portals'' and their dark-sector phenomenology warrant further exploration.

Another central challenge for this scenario is ensuring that the form of dark matter is consistent with self-interaction cross-section bounds, such as those famously arising from the Bullet Cluster observations. The main danger comes from having too much of the dark matter in the form of dark atoms made of dark electrons, dark protons and dark neutrons, with excessively large atomic cross-sections. Even though we considered the region of parameter space where the dark neutrons are lighter (and stable) and the dark protons are slightly heavier and unstable, it is possible for dark protons to be stabilized within dark nuclei if these form. This requires careful consideration of dark BBN. In this paper, we chose a simple and safe region of parameter space, in which dark protons decay before dark BBN can begin and where we know that dark dineutron bound states do not form, so that we can confidently deduce that dark matter takes the form of free dark neutrons. Their dark strong-interactions cross sections are well within acceptable bounds. However, there is a larger parameter space of more complex options when the dark proton is long-lived enough for dark deuterium to form and initiate dark BBN. Given that (subdominant) exotic dark atoms may be phenomenologically interesting (leading, for example, to cosmological consequences such as the imprint of dark acoustic oscillations in the matter power spectrum; see Refs.~\cite{Kaplan:2009de,Kaplan:2011yj,Cyr-Racine:2013fsa,Cyr-Racine:2012tfp,Bansal:2022qbi} among others), the dark BBN options are worth studying in greater detail.

Finally, there is another branch of dark sector models, namely those in which even a modest $\Ztwo$-breaking allows the new physics solving the dark hierarchy problem to Higgs dark electromagnetism, making the dark photon massive $\sim {\cal O}({\rm TeV})$. This dramatically changes the cosmological constraints and particle phenomenology. We plan to explore such models in future work.

\acknowledgments{

The authors thank Zackaria Chacko, Zohreh Davoudi, Junwu Huang, Saurabh Kadam, Gustavo Marques-Tavares, Rabindra Mohapatra, Shmuel Nussinov, Daniel Stolarski, and Michael Wagman for useful discussions. The authors would also like to thank the anonymous referee, whose suggestions helped improve this paper. MBA thanks Stephanie Buen Abad for proofreading this manuscript. The authors were supported by NSF grant PHY-2210361 and the Maryland Center for Fundamental Physics. In addition, AB was supported by DOE grant DE-SC-0013642 and by Fermi Research Alliance, LLC under Contract No. DE-AC02-07CH11359 with the U.S. Department of Energy, Office of Science, Office of High Energy Physics.
}

\bibliography{dm_baryon}
\bibliographystyle{apsrev4-1}

\end{document}